\documentclass[aps,prb,showpacs,twocolumn,superscriptaddress]{revtex4-1}
\usepackage{bm,color,amsmath,amssymb,mathrsfs,latexsym,graphicx,psfrag,accents,float}
\pdfoutput=1

\newcommand{\imp}{\;\;\Rightarrow\;\;}

\newcommand{\braket}[2]{\big\langle #1 \big| #2 \big\rangle}
\newcommand{\bra}[1]{\big\langle#1\big|}
\newcommand{\ket}[1]{\big|#1\big\rangle}

\newcommand{\bea}{\begin{eqnarray}}
\newcommand{\enea}{\end{eqnarray}}
\newcommand{\beq}{\begin{equation}}
\newcommand{\eneq}{\end{equation}}

\newcommand{\pdg}[1]{{#1}^{\phantom{\dagger}}}

\newcommand{\lin}{\notag \\}

\newcommand{\eq}{=&\;}
\newcommand{\ab}{\alpha\beta}

\newcommand{\low}{L$\ddot{\text{o}}$wdin\;}
\newcommand{\tex}[1]{\sma{\text{#1}}}

\newcommand{\calL}{{\cal L}}
\newcommand{\W}{{\cal W}}
\newcommand{\WL}{{\cal W}_{\sma{A}}}

\newcommand{\inv}{{\cal I}}

\newcommand{\bpm}{\begin{pmatrix}}
\newcommand{\epm}{\end{pmatrix}}

\newcommand{\bal}{\begin{align}}
\newcommand{\eal}{\end{align}}

\newcommand{\si}{\;\text{sin}\,}

\newcommand{\co}{\;\text{\text{cos}}\,}

\newcommand{\Np}{N_{\sma{(\texttt{+}1)}}}
\newcommand{\Nm}{N_{\sma{(\mo)}}}
\newcommand{\Ncc}{N_{\sma{(cc)}}}
\newcommand{\WKy}{\W_{\sma{K_y}}}
\newcommand{\Wky}{\W_{\sma{k_y}}}

\newcommand{\half}{\tfrac{1}{2}}

\newcommand{\Mp}{M_{\sma{(1)}}}
\newcommand{\Mm}{M_{\sma{(\mo)}}}
\newcommand{\Mcc}{M_{\sma{(cc)}}}

\newcommand{\dg}[1]{#1^{\scriptstyle{\dagger}}}

\newcommand{\sma}[1]{\scriptscriptstyle{#1}}

\newcommand{\kyi}{K_{{y}}}

\newcommand{\ki}{k^{\sma{\text{inv}}}}
\newcommand{\noc}{n_{\sma{{occ}}}}
\newcommand{\ntot}{n_{\sma{{tot}}}}

\newcommand{\chw}{{\cal D}}
\newcommand{\Z}{\mathbb{Z}}
\newcommand{\Zp}{\mathbb{Z}^{\sma{\geq}}}
\newcommand{\Rin}{ {\Nm} }
\newcommand{\mo}{\text{-}1}
\newcommand{\pone}{\texttt{+}1}

\newcommand{\noi}[1]{\noindent (#1)}

\newcommand{\qed}{\nobreak \ifvmode \relax \else
      \ifdim\lastskip<1.5em \hskip-\lastskip
      \hskip1.5em plus0em minus0.5em \fi \nobreak
      \vrule height0.75em width0.5em depth0.25em\fi}

\begin{document}
\title{Wilson-Loop Characterization of Inversion-Symmetric Topological Insulators}
 \author{A. Alexandradinata} \affiliation{Department of Physics, Princeton University, Princeton,
  NJ 08544} 
  \author{Xi Dai}
  \affiliation{Beijing National Laboratory for Condensed Matter Physics and Institute of Physics, Chinese Academy of Sciences, Beijing 100080, China}
  \author{B. Andrei Bernevig} \affiliation{Department of Physics, Princeton University, Princeton,
  NJ 08544}
  

\begin{abstract}
The ground state of translationally-invariant insulators comprise bands which can assume topologically distinct structures. There are few known examples where this distinction is enforced by a point-group symmetry \emph{alone}. In this paper we show that 1D and 2D insulators with the simplest point-group symmetry -- inversion -- have a $\Zp$ classification. In 2D, we identify a relative winding number that is solely protected by inversion symmetry. By analysis of Berry phases, we show that this invariant has similarities with the first Chern class (of time-reversal breaking insulators), but is more closely analogous to the $\Z_2$ invariant (of time-reversal invariant insulators). Implications of our work are discussed in holonomy, the geometric-phase theory of polarization, the theory of maximally-localized Wannier functions, and in the entanglement spectrum. 
\end{abstract}
\date{\today}

\pacs{74.20.Mn, 74.20.Rp, 74.25.Jb, 74.72.Jb}

\maketitle

\begin{figure}
\centering
\includegraphics[width=8.8 cm]{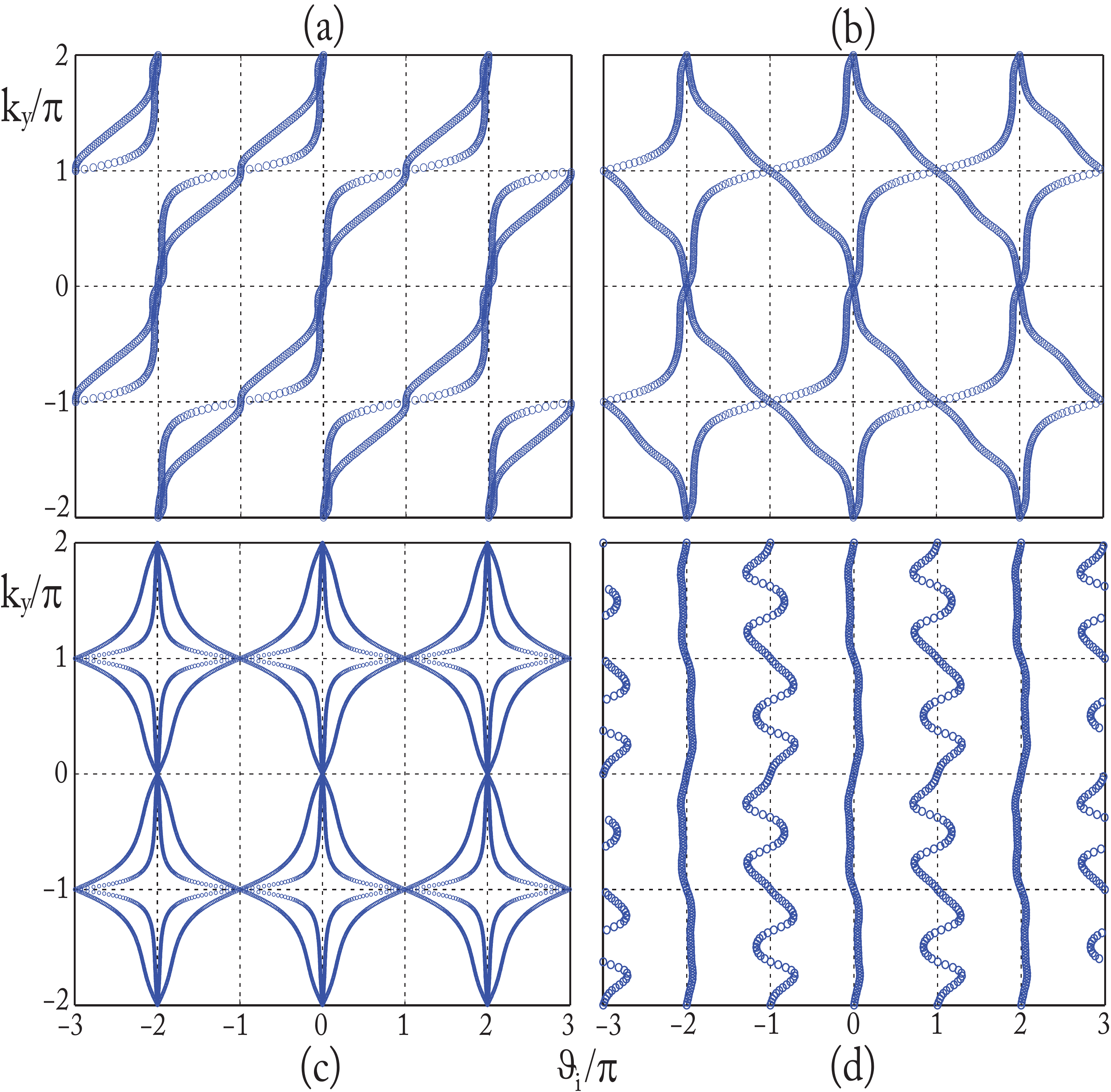}
\caption{ Non-Abelian Berry phases for four distinct insulators. Each figure contains three unit cells in an effective 1D-lattice along $\hat{x}$; we track the real-space trajectories of the phases (blue circles) in $\hat{x}$ as momentum $k_y$ is varied through two adiabatic cycles. (a) An insulator with Chern number $C_1=2$ and zero relative winding ($W$); this is modelled by the Hamiltonian (\ref{eq:modeltrs}) with parameters $m=3$ and $\delta=2$. (b) An insulator with relative winding $W=1$ and zero $C_1$; this is modelled by Eq. (\ref{eq:modeltrs}) with $m=3$ and $\delta=1$. (c) $W=2$ and $C_1=0$; this is modelled by Eq. (\ref{eq:modeltrs4band}). (d) $W=0$ and $C_1=0$; the insulator has a nontrivial polarization, and is modelled by Eq. (\ref{eq:model2occ2d}) with parameters $\alpha=\mo.5, \beta=1.5, \delta=\mo$. }\label{fig:windingnumberfigures}
\end{figure}

There is strong evidence supporting the view that the topological properties of a condensed-matter system are encoded in its ground state alone.\cite{wenbook, Hamma2004,levin2006, Kitaev2006,li2008,BrayAli2009,flammia2009, aa2011, Thomale2010A,Thomale2010B,Pollmann2010,fidkowski2010,Prodan2010, regnault2009, haldane2009,Kargarian2010, lauchli-10prl156404, lauchli-NJP-1367-2630, bergholtz-arXiv1006.3658B,PhysRevLett.106.100405,rodriguez-arXiv1007.5356R,papic-arXiv1008.5087P,hermanns-1009arXiv4199H, PhysRevB.80.201303, 2011arXiv1103.5437Q,2011arXiv1103.0772Z, PhysRevB.83.115322, thomale-2010arXiv1010.4837T, poilblanc-105prl077202, turner2011,  fidkowski-81prb134509, yao-105prl080501, 2010PhRvB..81f4439P, 2008PhRvA..78c2329C, PhysRevB.83.045110,2011arXiv1104.2544S, Poilblanc2011, 2010arXiv1002.2931F, 2011arXiv1104.1139H, Cirac2011, 2011arXiv1105.4808D, PhysRevA.83.013620, 2011arXiv1104.5157D, ryu2006,Sterdyniak2010} The ground state of translationally-invariant insulators comprise bands, which can assume topologically distinct structures. Bands are deemed distinct when they are not connected by continuous reparametrization of the Hamiltonian that preserves the energy gap. Some bands are distinct only because some reparametrizations are disallowed by a certain symmetry; in this sense we say that the topological distinction is protected by that symmetry. The symmetries which are ubiquitous in crystals are the point-group symmetries, which involve transformations that preserve a spatial point. Despite the large number of space groups in nature, there are few known examples in which the topological distinction is protected by a point-group symmetry \emph{alone}.\cite{Hsieh2012,Xu2012} In this paper we show that such distinction exists for insulators with arguably the simplest point-group symmetry -- inversion ($\inv$).\cite{hughes2011,turner2009,turner2010A,turner2012,fu2007a}\\

In search for a tool to identify topological structure in bands, we note that the description of translationally-invariant insulators has a local gauge redundancy -- its ground state is invariant under a unitary transformation in the subspace of occupied bands. Since all topological quantities must be invariant under this transformation, the natural objects to investigate are the Berry phase factors acquired around a loop, which are known to be gauge-invariant quantites.\cite{kane2005A,kane2005B,bernevig2006a,wilson1974,fidkowski2011,baskaran1988,ryu2010,yu2011,lee2008,schnyder2008A} We are proposing that distinct bands can be distinguished by holonomy, \emph{i.e.}, parallel transport through certain non-contractible loops in the Brillouin zone. Holonomies are known to have diverse applications in physics.\cite{leone2011,ekert2000,recati2002} The matrix representation of parallel transport is called a Wilson loop ($\W$), and its eigenspectrum comprise the non-Abelian Berry phase factors.\cite{wilczek1984,berry1984,zak1982,zak1989}  \\

A topological insulator cannot be continuously transformed to a direct-product state. This corresponds to a limit where all hoppings between atoms are turned off, so the ground state is a direct product of atomic wavefunctions. Such a limit is easily stated for a monatomic Bravais lattice: all band eigenfunctions are independent of crystal momentum, and parallel transport is trivial, \emph{i.e.},  the Wilson loop $\W$ equals the identity in the occupied subspace. Then a sufficient criterion for nontriviality is that a subset of the $\W$-eigenspectrum is robustly fixed to a value other than $\pone$. In this paper we demonstrate that some 1D $\inv$-symmetric insulators have a number ($\Rin$) of $\W$-eigenvalues that are symmetry-fixed to $\mo$. This number $\Rin \in \Zp$ classifies the 1D insulator; here, $\Zp$ denotes the set of nonnegative integers. $\Rin$ is completely determined by the symmetry representations of the occupied wavefunctions at inversion-invariant momenta -- momenta which satisfy $k=-k$ up to a reciprocal lattice vector. The even-parity (odd-parity) wavefunctions are defined to have inversion eigenvalues $+1$ ($-1$). If there are $n_{\sma{(-)}}(0)$ odd-parity wavefunctions at momentum $k=0$ and $n_{\sma{(-)}}(\pi)$ of them at $k=\pi$, we find that $\Rin$ quantifies a change in the group representations between $0$ and $\pi$:
\bal \label{eq:invariantintro}
\Rin = | n_{\sma{(-)}}(0) - n_{\sma{(-)}}(\pi)|.
\end{align}

In 1D, the inversion eigenvalues tell the whole story. Is this also true for the 2D insulator? No, we find there exists insulators with the same inversion eigenvalues, but with distinct band structures. A case in point is an $\inv$-symmetric insulator with Chern number $2$ -- it is a time-reversal breaking insulator which displays a quantum anomalous Hall effect.\cite{Haldane1988} In the geometric-phase theory of polarization, the non-Abelian Berry phases are identified as centers of charge, and they are functionally dependent on an adiabatic parameter.\cite{kingsmith1993} For a Chern insulator, these phases are known to have a \emph{center-of-mass} winding number, as plotted in Fig. \ref{fig:windingnumberfigures}-a; this implies a net transfer of charge in one adiabatic cycle. In Fig. \ref{fig:windingnumberfigures}-b, the Berry phases are strikingly different, yet they correspond to an insulator with the same inversion eigenvalues. The Berry phases come in pairs, and each member of a pair winds in a direction opposite to the other member. We identify this insulator as having a nontrivial \emph{relative} winding number $W$, which is protected only by inversion symmetry; $W$ provides a $\Zp$ classification of the 2D insulator. Both insulators of Fig. \ref{fig:windingnumberfigures}-a and -b share in common that their Berry phases interpolate across the maximal possible range $[0,2\pi)$ in one adiabatic cycle. Such a property, called spectral flow, is shared by all Chern insulators, and also the time-reversal invariant $\Z_2$ insulators.\cite{fu2006,yu2011,soluyanov2011} In this sense, $W$ is the inversion-analog of the first Chern class and the $\Z_2$ invariant.\\

Our findings are relevant to the theory of maximally-localized Wannier functions (MLWF). In 1D, Berry phases represent spatial coordinates of the MLWF. In higher dimensions, these phases are the spatial coordinates of hybrid WF's, which maximally localize along one direction, but extend in the remaining directions as a Bloch wave.\cite{kivelson1982b,marzari1997} The applications of MLWF's and their hybrid cousins are manifold: to name a few examples, they are used to analyze chemical bonding, and to locally probe electric polarization and orbital magnetization.\cite{marzari2012} In our paper, we derive a mapping between inversion eigenvalues and Berry phases, which strongly constrains the (hybrid) MLWF's of any inversion-symmetric insulator. Our work is an extension of Kohn's single-band result to insulators with interband degeneracies, such as those enforced by time-reversal symmetry.\cite{herring1937,kohn1959}\\

While we explicitly discuss insulators, our results may be generalized to any system, fermionic or bosonic, with discrete translational symmetry and an energy gap. Examples include photonic crystals with a bandgap, and cold atoms in an optical lattice. The dimension in `1D (2D) insulator' refers to that in momentum space. Thus, `1D (2D) insulator' may refer to a material of larger spatial dimension, but with a 1D (2D) Brillouin zone; in such cases `inversion symmetry' in 1D (2D) must be understood as a mirror ($C_2$ rotational) symmetry.  A case in point is a 3D optical lattice which is periodic only in $\hat{x}$, and is symmetric under $x \rightarrow -x$.\cite{atala2013} Moreover, our results also apply to 1D and 2D submanifolds embedded in larger-dimensional Brillouin zones. These are submanifolds which are mapped to themselves under mirror or $C_2$ rotation. For example, a plane of constant $k_z=0$ in a 3D $\inv$-symmetric insulator may possess a nontrivial relative winding.\\

The outline of this paper: in Sec. \ref{sec:intro}, we introduce Brillouin-zone Wilson loops and explain  their connections with polarization and holonomy. We analyze the 1D $\inv$-symmetric insulator in Sec. \ref{sec:invconstrain}. Here, we (i) derive a mapping between $\inv$ and $\W$ eigenvalues, (ii) formulate the topological index $\Rin \in \Zp$, and establish a connection between $\Rin$ and the entanglement spectrum. We analyze the 2D $\inv$-symmetric insulator in Sec. \ref{sec:2D}. Here, we (a) analyze  Chern insulators and $\Z_2$ insulators with inversion symmetry, (b) and identify a relative winding number $W \in \Zp$ which characterizes $\inv$-protected spectral flow. In Sec. \ref{app:conclusion}, we discuss the experimental implications. \\

\section{Introduction to Brillouin-zone Wilson loops} \label{sec:intro}

The Brillouin-zone Wilson loop is pedagogically introduced; we describe its role in holonomy (Sec. \ref{sec:holonomy}) and in the geometric-phase theory of polarization (Sec. \ref{sec:1dpolarization}). In Sec. \ref{sec:tightWilson}, we explain the construction of a coarse-grained Wilson loop from a tight-binding Hamiltonian, for the purpose of numerically computing topological invariants.\\

The Hamiltonian for a single particle is 
\bal \label{eq:genhamil}
H = \frac{p^2}{2m} + V(r).
\end{align}
If $V(r) = V(r+R)$ for any lattice vector $R$, this Hamiltonian is symmetric under discrete translations. Consequently, $H$ decouples into representations labelled by the crystal momentum $k$; an eigenstate in the $n$'th band may be written in Bloch form: $\psi^n_k(r) = e^{ikr}\,u^n_k(r)$, where $u^n_k(r) = u^n_k(r+R)$ is a function that is periodic in lattice translations, and also satisfies:
\bal \label{eq:blochhamil}
\bigg[\,\frac{(p+\hbar k)^2}{2m} + V(r)-\varepsilon^n_k\,\bigg]\;u^n_k(r) = 0.
\end{align}
Each eigenstate has a corresponding projection ${\cal P}^n_k(r,r') = u^n_k(r)\,u^{n*}_k(r')$; the many-body ground-state  is a single Slater determinant of all single-particle eigenstates with energies less than the Fermi energy. The topological properties of an insulator are invariant under transformations of the Hamiltonian that preserve both the energy gap and the symmetry that stabilizes the topological phase. We perform one such transformation by setting the energies of all eigenstates $u^n_k$ below (above) the Fermi energy equal to $\varepsilon_{\text{-}}$ ($\varepsilon_{\sma{+}}$). Denoting the projection onto the $\noc$ occupied bands as ${\cal P}^{\tex{occ}}_k = \sum_{i=1}^{\noc}\,{\cal P}^n_k$, we express the resultant flat-band Hamiltonian as
\bal \label{eq:flatband2}
{\cal H}_{\tex{F}}(k) =  (\varepsilon_{\text{-}} - \varepsilon_{\sma{+}}) \;{\cal P}^{\tex{occ}}_k + \varepsilon_{\sma{+}} I.
\end{align}
Eq. (\ref{eq:flatband2}) has a gauge redundancy which is not apparent in (\ref{eq:blochhamil}) -- the ground-state projection ${\cal P}^{\tex{occ}}_k$  is invariant under a local $U(\noc)$ gauge transformation in the $\noc$-dimensional subspace of occupied bands: $u^n_k \rightarrow u^m_k M^{mn}_k$ with $m,n = 1 \ldots \noc$ and $M_k^{\text{-}1} =\dg{M}_k$.

\subsection{Wilson Loop as arising from Holonomy} \label{sec:holonomy}

The adiabatic transport of a  ground state at initial momentum $k^{\sma{(i)}}$ to a final momentum $k$ involves a unitary rotation of the basis vectors $u^n_k$ in the subspace of occupied bands. This $U(\noc)$ rotation is affected by a Wilson-line matrix ${{\cal W}}_{k \leftarrow k^{\sma{(i)}}}$ that maps the subspace of occupied bands at $k^{\sma{(i)}}$ to the subspace of occupied bands at $k$. ${\cal W}$ is known to satisfy a parallel transport equation
\bal \label{eq:paralleltransport}
\tfrac{\partial}{\partial {k_{\mu}}} \,{\cal W}_{k \leftarrow  k^{\sma{(i)}}} \eq   \text{-}\, C_{\mu}(k) \;{\cal W}_{k \leftarrow k^{\sma{(i)}}},
\end{align}
with the Berry-Wilczek-Zee connection $C$ defined as\cite{wilczek1984,berry1984} 
\bal \label{eq:fullconn}
C^{mn}_{\mu}(k) = \int \,d^dr\;u^{m}_k(r)^*\,\tfrac{\partial}{\partial k_{\mu}}\, u^n_k(r).
\end{align}
Here, $k_{\mu}\in \{k_1,\ldots,k_d\}$ denote momenta components in a $d$-dimensional BZ. Eq. (\ref{eq:paralleltransport}) is pedagogically derived in App. \ref{app:paralleltransport}. This differential equation has the path-ordered solution
\bal \label{eq:Wilsonpathordered}
{\cal W}_{k \leftarrow  k^{\sma{(i)}}}({\cal L}) \eq \text{T}\; \text{exp}\,\big[\,{\text{-}\int_{\cal L} \;C_{\mu}(q)\; dq_{\mu}} \,\big]
\end{align}
for a path ${\cal L}$ that connects momenta $k$ and $k^{\sma{(i)}}$. If $k=k^{\sma{(i)}}$ modulo a reciprocal lattice vector, ${\cal L}$ forms a non-contractible loop in the BZ; we denote the resultant $U(\noc)$ Wilson loop as $\W$: 
\bal \label{eq:wilsonfull}
\W = \text{T} \;\text{exp}\,\big[\,\text{-}\int dk_{\mu}\;C_{\mu}(k)\,\big].
\end{align}
As Zak demonstrated for $\noc=1$, the Abelian $\W$ is nothing less than the Berry phase factor acquired by a Bloch wave around a cyclic evolution.\cite{zak1982,zak1989} For general $\noc$, (\ref{eq:wilsonfull}) forms a matrix representation of a holonomy, \emph{i.e.}, a parallel transport map.\cite{leone2011} The eigenvalues of this matrix are the non-Abelian Berry phase factors, which quantify a change relative to a periodic gauge, defined as $\psi^m_k = \psi^m_{k+2\pi}$; the phase factors are invariant under gauge transformations that preserve this periodic condition.

\subsection{Wilson Loops in the Geometric-Phase Theory of Polarization} \label{sec:1dpolarization}

Let us derive a well\text{-}known relation between $\W$ and the polarization of a 1D insulator. For a group of $\noc$ occupied bands, the delocalized Bloch waves $\psi^m_k$ form an orthonormal basis in the occupied Hilbert space; $m=1,2,\ldots,\noc$. Assuming a periodic gauge   ($\psi^m_k = \psi^m_{k+2\pi}$), we formulate the theory of polarization in an alternative basis of localized Wannier functions (WF):
\bal \label{eq:generalizedwann2}
\Psi^{(j)}(x-R) = \int \frac{dk}{2\pi} \,e^{-ikR}\,\sum_{n=1}^{\noc}\,O(k)_{jn}\;\psi_{k}^n(x).
\end{align}
Each WF $\Psi^{(j)}(x-R)$ is labelled by the unit cell $R \in \mathbb{Z}$ and an index $j =1,2,\,\ldots \,\noc$. If $\noc =1$, $O(k)$ is a momentum-dependent phase; if $\noc>1$, $O(k)$ is a $U(\noc)$ matrix that affects rotations in the subspace of degenerate bands. The gauge freedom in $O(k)$ is fixed, up to trivial $U(1)$ phase windings, by requiring that the WF's are maximally localized.\cite{marzari1997} Equivalently, we require that the WF's are eigenfunctions of the projected position operator ${\cal P}^{\tex{occ}}\,\hat{x}\,{\cal P}^{\tex{occ}}$:\cite{kivelson1982b}
\bal \label{eq:pxpeig}
\big(\;{\cal P}^{\tex{occ}}\,\hat{x}\,{\cal P}^{\tex{occ}} - \tfrac{1}{2\pi}\,\vartheta^{(j)} - R\;\big)\;\Psi^{(j)}(x-R) = 0.
\end{align}
Here,  ${\cal P}^{\tex{occ}}$ projects to the occupied subspace, and a unit length separates two unit cells. We show in App. \ref{app:projectedposition} that the spectrum $\{\vartheta^{(j)}\}$ coincide with the phases of the $\W$-spectrum.  Through (\ref{eq:pxpeig}) we map the electron density to classical point charges at the positions $\{\vartheta^{(j)}/2\pi + R\}$, for all $R$; we call these positions the Wannier centers. $\vartheta=\pi$ corresponds to a Wannier center that is displaced from the origin by half a unit length. We interchangeably use `Wannier center' and `Berry phase' to mean the same quantity.

\subsection{The Tight-Binding Wilson Loop} \label{sec:tightWilson}

As defined in (\ref{eq:wilsonfull}), the continuum Wilson loop $\W$ is expensive to compute numerically. In this Section, we introduce the tight-binding Wilson loop $\WL$, which is computable with relative ease. In the tight-binding approximation, we restrict our attention to nearly-degenerate orbitals which diagonalize the atomic Hamiltonian. The Hamiltonian of (\ref{eq:genhamil}) reduces to the variational form 
\bal \label{eq:physH}
H = \sum_{k} c_{k \alpha}^\dagger \;[h(k)]_{\alpha \beta}\;  \pdg{c}_{k \beta}
\end{align}
with orbital indices $\alpha,\beta$; $[h(k)]_{\alpha \beta}$ are the matrix elements of (\ref{eq:genhamil}) in the Bloch basis of \low orbitals.\cite{slater1954,goringe1997,lowdin1950} In (\ref{eq:physH}) and the rest of the paper, we sum over repeated indices. Let us denote the $j$-th normal mode operator as $\dg{\gamma}_{j k} = [U^{j}_{k}]_{\beta} \,\dg{c}_{k \beta}$; in bra\text{-}ket notation, the corresponding projection is $ {\cal P}^j_k =  | U^{j}_k \rangle  \langle U^{j}_k| = \dg{\gamma}_{j k} | 0 \rangle  \langle 0|\pdg{\gamma}_{j k}$. We define the tight-binding connection $A$ as
\bal \label{eq:tightconn}
A_{\mu}^{mn}(k) = \bra{U^m_k}\,\tfrac{\partial}{\partial k_{\mu}}\,\ket{U^n_k},
\end{align}
The tight-binding Wilson loop $\WL$ is defined as the path-ordered exponential of the tight-binding connection $A$:
\bal \label{eq:Wilsonloopdefined}
{\WL} \eq \text{T}\; \text{exp}\,\big[\,{\text{-}\int dq_{\mu}\;A_{\mu}(q) }\,\big].
\end{align}
A discretized expression of $\WL$ is obtained by dividing a BZ loop $\cal{L}$ into infinitesimally-separated momenta: $\{k^{\sma{(0)}}+G,k^{\sma{(N)}},k^{\sma{(N\text{-}1)}} \ldots k^{\sma{(2)}}, k^{\sma{(1)}},k^{\sma{(0)}}\}$ with $N \gg 1$. Defining ${\cal P}^{\tex{occ}}_k = \sum_{j=1}^{\noc} \,{\cal P}^j_k$ as the projection to the occupied bands, $\WL$ may be expressed as a path-ordered product of projections, sandwiched by tight-binding eigenfunctions at the base and end points: 
\bal \label{eq:wilsonprojection}
[\WL({\calL})]^{mn} = \bra{U^m_{k^{\sma{(0)}}+G}} \; {\displaystyle \prod_{\alpha}^{k^{\sma{(0)}}+G \leftarrow k^{\sma{(0)}}}}\, {\cal P}^{\tex{occ}}_{k^{\sma{(\alpha)}}}\; \ket{U^n_{k^{\sma{(0)}}}} .
\end{align}
The product of projections are path-ordered along ${\cal L}$, with the earlier-time momenta positioned to the right. The tight-binding Wilson loop (\ref{eq:Wilsonloopdefined}) and the continuum Wilson loop (\ref{eq:wilsonfull}) generically have different eigenspectra; see App. \ref{app:connections}. However, we show in App. \ref{app:1dconstraint} that both Wilson loops are identically constrained by inversion symmetry, and consequently their spectra are nearly identical. In particular, the topological index $\Rin$, as defined in the Introduction, may be extracted from either Wilson loop. To simplify the presentation in the next Section, we speak only of the tight-binding Wilson loop, which we henceforth denote as $\W$.

\section{Wilson-loop characterization of the 1D Inversion-symmetric insulator } \label{sec:invconstrain}

We highlight distinctive features of the $\inv$-symmetric Wilson loop (Sec. \ref{sec:symmconst1D}), and present a mapping between the $\W$-eigenvalues and the $\inv$ eigenvalues of the ground state (Sec. \ref{sec:theorem1d}). In Sec. \ref{sec:topinv}, we formulate  the  topological index $\Rin\in \Zp$ that classifies $\inv$-symmetric insulators. In addition, we relate $\Rin$  to a well-known $\mathbb{Z}_2$ index that distinguishes the electric responses of these insulators. 

\subsection{Constraints on the Wilson Loop due to Inversion Symmetry } \label{sec:symmconst1D} 

In 1D, inversion  maps the spatial coordinate $x \rightarrow \text{-}x$; we choose the center of inversion as the spatial origin ($x=0$). We define the unit cell such that the unit cell enclosing the spatial origin is mapped to itself under inversion. At inversion-invariant momenta ($0$ and $\pi$), the wavefunctions transform in irreducible representations of inversion -- even-parity (odd-parity) wavefunctions are defined to have inversion eigenvalues $+1$ ($-1$). Due to the discrete translational symmetry, an inversion center at the origin implies existence of inversion centers at $x=1/2,1,3/2,2 \ldots$, in units where the distance between two unit cells is unity. For any integer $n$, we call $x =n$ ($x=n+1/2$) a primary (secondary) site.\\

As shown in App. \ref{app:connections}, a tight-binding Hamiltonian with $\inv$ symmetry satisfies $\wp \,h(k)\,\wp = h(\text{-}k)$, where $\wp$ is the representation of inversion in the basis of \low orbitals. This symmetry implies that the \emph{set} of $\W$-eigenvalues is equal to its complex conjugate. Equivalently, the eigenvalues of $\W$ are constrained to $\pm 1$ or otherwise form complex-conjugate pairs. Such a constraint may be intuited from the theory of MLWF's: the MLWF's  (i) are centered at the primary site ($\pone$) or (ii) at the secondary site ($\text{-}1$) or (iii) form pairs that center equidistantly on opposite sides of a primary site ($\lambda \lambda^*$). In all three cases, the periodic configuration of Wannier centers is invariant under a spatial inversion $x \rightarrow \text{-}x$. The derivation of this constraint is left to App. \ref{app:1dconstraint}.

\subsection{1D: Mapping between Wilson-loop and Inversion Eigenvalues} \label{sec:theorem1d}

\noindent \emph{Definition:} For the occupied bands of an insulating phase, let us define the number of even- and odd-parity wavefunctions at $\ki=\{0,\pi\}$, as $n_{\sma{ (+)}}(\ki)$ and $n_{\sma{ (-)}}(\ki)$ respectively. Given this set of four numbers $\{ \,n_{(+)}(0), n_{(-)}(0), n_{(+)}(\pi), n_{(-)}(\pi)\, \}$, we identify the smallest of the four and label it as $n_s$, \emph{i.e.}, $n_s$ counts the fewest bands of one parity (FBOP) among both symmetric momentum $\ki$. $n_s=0$ if all the occupied wavefunctions at $k=0$ (or $\pi$) have the same parity. We label the momentum where FBOP lies as $k_s$ and the $\inv$ eigenvalue of FBOP as $\xi_s$. Let us identify the FBOP for the following examples.

\noindent (a) Consider a two-band insulator with $\inv$ eigenvalues $(++)$ at $k=0$ and $(+-)$ at $k=\pi$. The FBOP are the negative-$\inv$ bands  at $k_s=0$. None exists, so $n_s=0$. 

\noindent (b) Suppose we had a four-band insulator with $\inv$ eigenvalues $(++--)$ at $k=0$ and $(+++-)$ at $k=\pi$, the FBOP is the single negative-$\inv$ band at $k_s=\pi$, so $\xi_s=\text{-}1$ and $n_s=1$.

\noindent (c) If a subset of  the four numbers $\{ n_{\sma{ (\pm)}}(\ki)\}$ are equally small, we may denote any number in this subset as $n_s$. In a two-band example, we may encounter $n_{\sma{(+)}}(0) = n_{\sma{(-)}}(0) =n_{\sma{(+)}}(\pi)=n_{\sma{(-)}}(\pi)=1$. Then, one may label any of the four possibilities as the FBOP. \\

\noindent \emph{Mapping:} Given an inversion-symmetric insulator that is characterized by the quantities  $\{ \,n_{\sma{ (\pm)}}(\ki),n_s,k_s,\xi_s\, \}$, its Wilson-loop eigenspectrum consists of: 

\noindent(i) $(n_{\sma{ (+)}}(k_s+\pi) - n_s)$ number of $\text{-}\xi_s$ eigenvalues,

\noindent(ii) $(n_{\sma{ (-)}}(k_s+\pi) - n_s)$ number of $\xi_s$ eigenvalues, and

\noindent(iii) $n_s$ pairs of complex-conjugate eigenvalues.\\

\noindent In the above examples, the $\W$-spectrum of insulator (a) comprises one $\pone$ and one $\text{-}1$ eigenvalue; for insulator (b), there are one $\pone$ eigenvalue,  one $\text{-}1$ eigenvalue, and one complex-conjugate pair; insulator (c) has one complex-conjugate pair only. The proof of this mapping is detailed in App. \ref{app:1d4occbands}. The interested reader also may refer to App. \ref{app:casestudy}, where we undertake the case studies of the one- and two-band $\W$'s; these case studies offer an intuitive understanding of the above mapping, and also provide an alternate derivation that is specific to one and two occupied bands. For an insulator with one, two and  four occupied bands, we tabulate the possible  mappings in Tab. \ref{table1d1band}, \ref{table1d2band} and  \ref{table1d4band} respectively. We have described a mapping from $\inv$- to $\W$-eigenvalues; the reverse mapping is possible up to some arbitrariness. This is because the $\W$-spectrum is sensitive only to changes in the representations between $0$ and $\pi$, and is invariant under: (i) multiplication of all $\inv$-eigenvalues by a common factor $\text{-}1$, and (ii) interchanging the $\inv$-eigenvalues at $0$ with those at $\pi$.  \\

Let us define a $\Zp$ index, $\Rin$, as the number of $\text{-}1$ eigenvalues in the $\W$-spectrum. It is possible that one or more pairs of complex-conjugate eigenvalues are accidentally degenerate at $\text{-}1$; we exclude them from the definition of $\Rin$. Through the above mapping, we deduce that $\Rin$ is the absolute difference in the number of same-symmetry bands between $0$ and $\pi$, and quantifies the change in the group representation; \emph{cf.} Eq. (\ref{eq:invariantintro}). In the following Section, we argue that a nonzero $\Rin$ is an indication of topological nontriviality.  

\begin{table}[h]
	\centering
		\begin{tabular} {|c|c|} \hline
			$\inv$ eigenvalues  & $\W$-spectrum \\ \hline \hline
			$\{(+)\;(+)\}$ & $[+]$ \\ \hline
			$\{(+)\;(-)\}$ & $[-]$ \\ \hline
			\end{tabular}
		\caption{For an insulator with one occupied band, we tabulate the $\inv$ eigenvalues of the occupied band  at symmetric momenta $\{0,\pi\}$ and the corresponding $\W$-spectrum.  $+ \;(-)$ refers to an eigenvalue of $\pone \; (\text{-}1)$. $\{(+)\,(-)\}$ may refer to either (i) a positive-$\inv$ band at $k=0$, with a negative-$\inv$ band at $\pi$, or (ii) a negative-$\inv$ band at $0$, with a positive-$\inv$ band at $\pi$.  If two sets of $\inv$ eigenvalues (from two distinct insulators) are related by a global change in sign, they are mapped to the same $\W$ eigenvalue. For example, both $\{(+)\,(+)\}$ and $\{(-)\,(-)\}$ are mapped to $\W =\pone$. \label{table1d1band}}
\end{table}

\begin{table}[h]
	\centering
		\begin{tabular} {c|c|c|} \cline{2-3}
			& $\inv$ eigenvalues  & $\W$-spectrum \\ \hline \hline 
		 \multicolumn{1}{|c|}{(i)} & $\{(+ +)\;(+ +)\}$ & $[+ +]$ \\ \hline
			\multicolumn{1}{|c|}{(ii)} & $\{(+ +)\;(+ -)\}$ & $[+ -]$ \\ \hline
			\multicolumn{1}{|c|}{(iii)} & $\{(+ +)\;(- -)\}$ & $[- -]$ \\ \hline
			\multicolumn{1}{|c|}{(iv)} & $\{(+ -)\;(+ -)\}$ & $[\lambda \lambda^{\ast}]$ \\ \hline
		\end{tabular}
		\caption{For an insulator with two occupied bands, we tabulate the $\inv$ eigenvalues of the occupied bands at symmetric momenta $\{0,\pi\}$ and the corresponding $\W$-spectrum. We collect the $\inv$ eigenvalues $\xi_i$ at each symmetric momenta into $(\xi_1 \;\xi_2)$. \label{table1d2band}}
\end{table}

\begin{table}[h]
	\centering
		\begin{tabular} {|c|c|c|} \hline
			$\inv$ eigenvalues at $\{0,\pi\}$ & $\W$-spectrum\\ \hline \hline
			
			$\;\;\;\{(+ + + +)\;(+ + + +)\}\;\;\;$ & $[+ + + +]$ \\ \hline
			$\;\;\;\{(+ + + +)\;(+ + + -)\}\;\;\;$ & $[+ + + -]$ \\ \hline
			$\;\;\;\{(+ + + +)\;(+ + - -)\}\;\;\;$ & $[+ + - -]$ \\ \hline
			$\;\;\;\{(+ + + +)\;(+ - - -)\}\;\;\;$ & $[+ - - -]$ \\ \hline
			$\;\;\;\{(+ + + +)\;(- - - -)\}\;\;\;$ & $[- - - -]$ \\ \hline
			$\;\;\;\{(+ + + -)\;(+ + + -)\}\;\;\;$ & $[\lambda \lambda^{\ast} + +]$ \\ \hline
			$\;\;\;\{(+ + + -)\;(+ + - -)\}\;\;\;$ & $[\lambda \lambda^{\ast} + -]$  \\ \hline
			$\;\;\;\{(+ + + -)\;(+ - - -)\}\;\;\;$ & $[\lambda \lambda^{\ast} - -]$  \\ \hline
			$\;\;\;\{(+ + - -)\;(+ + - -)\}\;\;\;$ & $[\lambda \lambda^{\ast} \; \mu \mu^{\ast}]$ 	\\ \hline		\end{tabular}
		\caption{ For an insulator with four occupied bands, we tabulate the $\inv$ eigenvalues of the occupied bands at symmetric momenta $\{0,\pi\}$ and the corresponding $\W$-spectrum. We collect the $\inv$ eigenvalues $\xi_i$ at each symmetric momenta into $(\xi_1 \,\xi_2\,\xi_3\,\xi_4)$. 
		 \label{table1d4band}}
\end{table}

\subsection{1D: Topological Invariants from the Wilson Loop} \label{sec:topinv}   

\subsubsection{$\Zp$ Index: Number of Robust $\text{-}1$ Wilson-Loop Eigenvalues} \label{sec:zindex}

A topological insulator cannot be continuously transformed to a direct product of atomic wavefunctions. For a monatomic Bravais lattice, the wavefunctions of a direct-product state are independent of crystal momentum. We deduce from (\ref{eq:wilsonfull}) that $\W$ equals the identity, or that the MLWF's are centered at the primary site, where the atom lies -- we call this the atomic limit. A nonzero $\Rin$ index obstructs $\W$ from being tuned to the identity; this is a sufficient condition for nontriviality.  While complex-conjugate $\W$-eigenvalues $[\lambda, \lambda^{\ast}]$ also deviate from the atomic value of $\pone$, they are not a sufficient condition for nontriviality. The exact value of $\lambda$ is not fixed by symmetry; this arbitrariness reflects the range in this equivalence class, \emph{i.e.}, it is possible to tune the Hamiltonian and sweep the interval of allowed $\lambda$ while preserving both gap and symmetry. In particular, we can always tune the complex-conjugate eigenvalues to the trivial limit of $[\pone,\pone]$. It is possible that pairs of complex-conjugate eigenvalues are {accidentally} degenerate at $\text{-}1$. Unlike the $\Rin$ values of $\text{-}1$, these extra degeneracies are not protected by $\inv$ -- they generically destabilize under a soft deformation of the ground-state, hence they do not indicate a nontrivial phase. Hence, symmetry dictates that $\Nm$ is the \emph{minimum} number of $\mo$ eigenvalues; there are two implications:

\noi{i} Suppose we begin with a nontrivial insulator with $\Rin>0$, and we would like to transform it to an atomic insulator while preserving $\inv$ symmetry. The energy gap must close a minimum of $\Rin$ times at a symmetric momentum, before the atomic limit $(\Rin=0)$ is reached. In each gap-closing event, a pair of opposite-sign $\inv$-eigenvalues are inverted between occupied and unoccupied bands, thus reducing $\Rin$ by one. 

\noi{ii} Let us establish a connection between between holonomy and entanglement. Turner \emph{et al.} have demonstrated that 1D $\inv$-symmetric insulators manifest modes that robustly localize at boundaries created by a spatial entanglement cut.\cite{turner2009,turner2010A} As demonstrated in Ref. \onlinecite{hughes2011} and \onlinecite{turner2012}, the number ($\chi \in \Zp$) of stable modes at each entanglement boundary is equal to the absolute difference in the number of same-symmetry bands between $0$ and $\pi$. Thus, we identify $\chi$ with the $\W$-index $\Rin$, and show that the two formulations of nontriviality are equivalent. The presence of these mid-gap modes ensure that the entanglement entropy can never be tuned to zero by any gap- and symmetry-preserving transformation -- this has been proposed as a criterion for nontriviality that unifies all known topological insulators.\cite{hughes2011,chen2013} \\

The story is not so different with multiple atoms per unit cell, which constitute a molecule. In the limit of isolated molecules, inversion symmetry within a molecule imposes that the MLWF's are symmetric about the primary site (which is not necessarily where an atom sits). That is, each Wannier center either lies on the primary site, or form equidistant pairs on either side of it. Equivalently, the allowed $\W$-eigenvalues of a direct-product state are $1$ and $\{\lambda,\lambda^*\}$. If $\Nm>0$, this indicates a Wannier center at the secondary site, which can only arise from nontrivial interactions between molecules.

\subsubsection{$\Z_2$ Polarization Index: Determinant of $\W$}

The determinant of $\W$, which we define as $\chw$, is the exponentiated polarization of the 1D insulator. Since all $\W$-eigenvalues are either $\pm 1$ or form complex-conjugate pairs, $\chw$ is quantized to $\pm 1$ - the classification of the electric response is $\Z_2$, as is recognized in works such as Ref. \onlinecite{zak1989,hughes2011}. Moreover, $\chw$ is only determined by the number of $\text{-}1$ eigenvalues: $\chw = (\text{-}1)^{\Nm}$. Let us relate $\chw$ to the $\inv$ eigenvalues of the ground state. From (\ref{eq:invariantintro}) we have that
\bal \label{eq:14}
\chw \eq (\text{-}1)^{| n_{\sma{(-)}}(0) - n_{\sma{(-)}}(\pi)|}=\prod_{\ki=0,\pi} \;\prod_{m=1}^{\noc}\; \xi^{m}_{\ki},
\end{align}
where $\xi^{m}_{\ki}$ is the $\inv$ eigenvalue of the $m$'th band at symmetric momentum $\ki$. This concludes our discussion for 1D.

\section{Wilson-loop characterization of the 2D Inversion-Symmetric Insulator} \label{sec:2D}

The Wilson loop $\W$ is known to encode the first Chern class $C_1$; we present a summary of this relation in Sec. \ref{sec:Wilsonandpolarization}. 
In Sec. \ref{sec:symmconst2D}, we impose $\inv$ symmetry and investigate how the symmetry constrains $\W$ and the allowed Chern numbers. $\W$ is further  constrained if the insulator is also time-reversal symmetric (TRS) -- this is explored in Sec. \ref{sec:invtrs}. In Sec. \ref{sec:invspec}, we introduce a relative winding number $W$ that characterizes insulators with $\inv$-protected spectral flow.

\begin{figure}
\centering
\includegraphics[width=8.5cm]{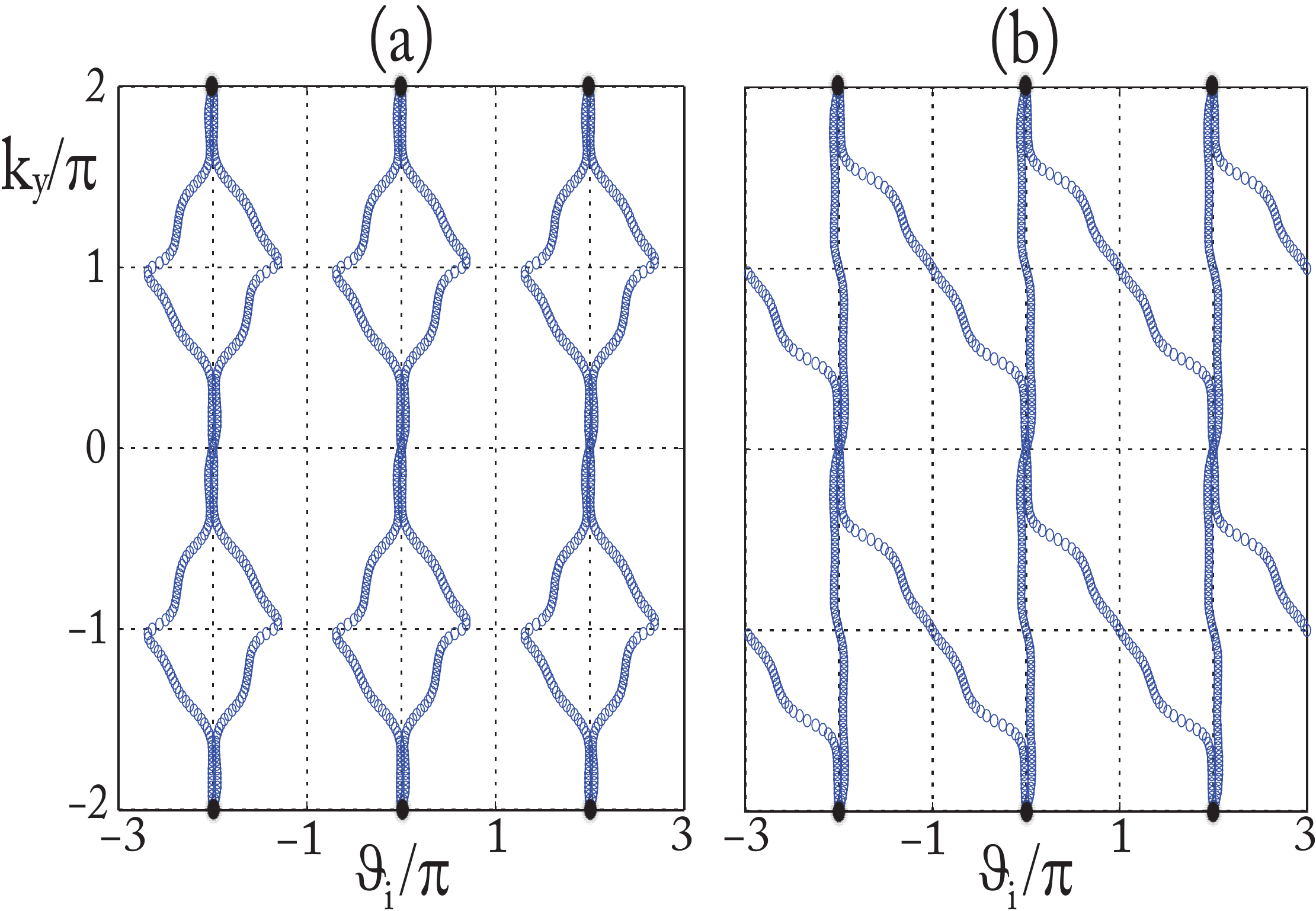}
\caption{The Wannier center flows for a 2D insulator with $\inv$ symmetry. Each figure contains three unit cells in an effective 1D-lattice along $\hat{x}$; the positions of the primary sites are indicated by black dots. The ground state contains two occupied bands, hence there are two Wannier centers in each unit cell. (a) This phase is realized in the model Hamiltonian (\ref{eq:model2occ2d}), with parameters $\alpha = \text{-}1.5$, $\beta=1.5$ and $\delta=1$. Upon $k_y \rightarrow k_y+2\pi$, the pair of Wannier centers in each unit cell exchange positions. There is no net transfer of charge between unit cells, hence $C_1=0$. (b) This phase is realized in the same model with $\alpha = \text{-}1.5$ and $\beta=0$. Upon $k_y \rightarrow k_y+2\pi$, \emph{one} Wannier center adiabatically flows to its neighboring unit cell on the left. This represents a quantized Hall current, hence $C_1=\text{-}1$. }\label{fig:pppm}
\end{figure}

\subsection{Wilson Loops and the First Chern Class} \label{sec:Wilsonandpolarization}

A well\text{-}known relation exists between Wilson loops and the integer quantum Hall effect. Let us apply an electric field along $\hat{y}$, which adiabatically translates all single-particle states in the parameter space $k_y$. To probe for a quantum Hall response, we study the $k_y$-evolution of the ground-state polarization (in $\hat{x}$) -- we are interested in Brillouin-zone $\W$'s at constant $k_y$: $\Wky = \text{T}\, \text{exp}\,({\text{-}\int dk_{x}\;C_{x}(k) })$. Let us denote the $\noc$ eigenvalues of $\Wky$ by the set $\{\text{exp}\,{(i\vartheta^m_{k_y})}\}$. The geometric phase ${\vartheta^m_{k_y}}/{2\pi}$ represents the center of a hybrid Wannier function (WF), which extends in $\hat{y}$ in the manner of a 1D Bloch wave, but localizes in $\hat{x}$ as a 1D WF; we refer to these phases as the Wannier centers.\cite{kingsmith1993,vanderbilt1993,resta1994,souza2004,mele2002,yaschenko1998,sipe1999} These hybrid WF's are similar to classical line charges; the quantum-mechanical electron density may be represented by a lattice of line charges. The derivative of the geometric phase, ${\accentset{\sma{\bullet}}{\vartheta}^m_{k_y}} \equiv d\, \vartheta^m/{d\, k_y}$, is interpreted as the real-space velocity (in $\hat{x}$) of the $m$'th Wannier center at time $k_y$. By integrating the velocities of all Wannier centers over a period $2\pi$, we obtain the net quantum Hall current.  Thus, we identify the first Chern class as the center-of-mass winding:\cite{thouless1982,qi2006,ProdanJMP2009,coh2009,BELLISSARD1994xj,Niu1985,gurarie2011}
\bal \label{eq:Chernpolarization}
C_1 \eq   \sum_{m=1}^{n_{\text{occ}}} \int \,{\accentset{\sma{\bullet}}{\vartheta}^m_{k_y}}\;\frac{dk_y}{2\pi}.
\end{align}
As we have shown in Sec. \ref{sec:1dpolarization}, polarization is directly related to the continuum Wilson loop, which is defined in (\ref{eq:wilsonfull}). However, the winding number in the tight-binding Wilson loop, as defined in Sec. \ref{sec:tightWilson}, is identical to that in  the continuum Wilson loop. This follows because their connections differ by an operator that is periodic in $k_y$ (\emph{cf.} Eq. (\ref{eq:positionop}) ). For the purpose of computing Chern numbers, both Wilson loops give identical results.\\

For illustration, we consider the 4-band model
\bal \label{eq:model2occ2d}
&h(k) =  \tfrac{1}{2}\,\Gamma_{13} +\tfrac{\alpha}{2} \,(\text{cos}\, k_x + \text{cos}\,k_y)\,(\Gamma_{30}+\Gamma_{03}) \lin
&+ (\Gamma_{12}+\Gamma_{31}) \si k_x   + (\Gamma_{21}+\Gamma_{32}) \,\si k_y  -\Gamma_{03} \lin
&+ \tfrac{\beta}{2} \,(\text{cos}\, k_x) \,(\text{cos}\, k_y-\delta)\, (\Gamma_{03}-\Gamma_{30}) ,
\end{align}
where $h(k)$ is a matrix in the tight-binding basis; \emph{cf.} (\ref{eq:physH}). $\Gamma_{ij}$ are defined as $\sigma_i \otimes \tau_j$; $\sigma_0$ ($\tau_0$) is the identity in spin (orbital) space; $\sigma_{i=1,2,3}$ ($\tau_{i=1,2,3}$) are Pauli matrices in spin (orbital) space.  The Hamiltonian possesses an $\inv$ symmetry: $\Gamma_{03} \,h(k)\,\Gamma_{03} = h(\text{-}k)$. The Fermi energy is chosen so that there are two occupied bands in the ground state. We tabulate the $\inv$ and $\W$-eigenvalues for various choices of the parameters ($\alpha, \beta$ and $\delta$) in Tab. \ref{tab:model2occ2d}. In Fig. \ref{fig:pppm}-a  we plot the $\Wky$-spectrum for a trivial insulator $(\alpha=\mo .5,\beta=1.5, \delta=1)$; Fig. \ref{fig:pppm}-b corresponds to a nontrivial insulator with $C_1=\mo$ $(\alpha=\mo .5,\beta=0, \delta=0)$.

\begin{table}[h]
	\centering
		\begin{tabular} {|c|c|c|c|c|c|c|c|c|} \hline
		\multicolumn{3}{|c|}{Parameters} & \multicolumn{4}{|c|}{$\inv$ eigenvalues} & \multicolumn{2}{|c|}{$\W_{\kyi}$ eigenvalues}   \\ \hline
		$\;\alpha\;$ & $\;\beta\;$ & $\;\delta\;$ &	$(0,0)$  & $ (\pi,0)$  & $ (0,\pi)$  & $ (\pi,\pi)$ & $\;\;\kyi=0\;\; $ & $ \kyi=\pi $  \\  \hline \hline
		$\text{-}1.5$ & $0$ & $0$ &  $(+ +)$ & $(+ +)$ & $(+ +)$ & $(+ -)$ & $[+ +]$ & $[+ -]$  \\ \hline
		$\text{-}1.5$ & $1.5$ & $1$ &	$(+ +)$ & $(+ +)$ & $(+ -)$ & $(+ -)$ & $[+ +]$ & $[\lambda  \lambda^*]$   \\ \hline
		$\text{-}1.5$ & $1.5$ & $\mo$ &	$(+ +)$ & $(+ -)$ & $(+ +)$ & $(+ -)$ & $[+ -]$ & $[+  -]$   \\ \hline
		\end{tabular}
		\caption{The $\inv$ and $\W_{\kyi}$ eigenvalues of the ground state of Hamiltonian (\ref{eq:model2occ2d}), for various parameters. \label{tab:model2occ2d}}
\end{table}

\subsection{The Inversion-Symmetric Wilson Loop and the Integer Quantum Hall Effect} \label{sec:symmconst2D}

Let us investigate the spectrum of the $\inv$-symmetric $\Wky$. As derived in App. \ref{app:2dconstraint}, we find that $\Wky$ is equivalent to the Hermitian adjoint of $\W_{\sma{\text{-}k_y}}$  by a unitary transformation, \emph{i.e.}, the \emph{sets} of $\W$-eigenvalues at $\pm k_y$ are equal up to complex conjugation: 
\bal \label{eq:equalityset2d}
\big\{ \text{exp}\,{i{\vartheta}_{\sma{k_y}}}  \big\} = \big\{ \text{exp}\,{\text{-}i{\vartheta}_{\text{-}k_y}} \big\}, 
\end{align}
as may be verified in Fig. \ref{fig:pppm}. At $\kyi \in \{0,\pi\}$, the 1D line of states behaves like a 1D $\inv$-symmetric insulator in two respects: (i) the eigenvalues of $\W_{\kyi}$ are constrained to $\pm 1$ or otherwise form complex-conjugate pairs. (ii) In 1D, the $\inv$ eigenvalues at $k=0$ and $\pi$ are related to $\W$-eigenvalues through the mapping of Sec. \ref{sec:theorem1d}; in 2D, the $\inv$ eigenvalues at momenta $(0,K_y)$ and $(\pi,K_y)$ are related to the eigenvalues of $\W_{K_y}$ through the same mapping. \\

Let us define the number of robust $\mo$ eigenvalues in the spectra of $\W_{0}$ and $\W_{\pi}$ as $\Nm(0)$ and $\Nm(\pi)$ respectively. During the adiabatic evolution, $\Rin(K_y)$ is the number of Wannier centers that localize at each secondary site at time $\kyi$. A difference in the indices $\Nm(0)$ and $\Nm(\pi)$ implies a net Hall current; moreover, the parity of the Chern number is determined through
\bal \label{eq:w2flow}
\Rin(0) - \Rin(\pi) = C_1 \;\;\text{mod}\;\;2.
\end{align}
The two parities of $C_1$ correspond to the following situations:

\noindent (i) Suppose the parities of $\Nm(\kyi)$ differ. Between $\kyi=0$ and $\pi$, an \emph{odd} number of Wannier centers	 must interpolate between the secondary sites ($\W$-eigenvalue of $\text{-}1$) and the non-secondary sites, which include  (a) the primary sites ($\pone$) (see Fig. \ref{fig:pppm}-b), and (b) the complex-conjugate sites ($\lambda \,\lambda^*$).  The {net} translation of Wannier centers in the interval $k_y \in [0,\pi]$ is \emph{half} an odd integer. It follows from (\ref{eq:Chernpolarization}) and (\ref{eq:equalityset2d}) that $C_1$ is odd. 

\noindent (ii) An analogous argument emerges when the  parities of $\Nm(\kyi)$ are equal. Now an \emph{even} number of 	Wannier centers must interpolate between the secondary sites (at time $\kyi=0$) and the non-secondary sites (at time $\kyi=\pi$). One possible scenario is illustrated in Fig. (\ref{fig:pppm}-a). The conclusion is that $C_1$ is even. 

\noindent It follows from Eq. (\ref{eq:w2flow}), (\ref{eq:invariantintro}) and (\ref{eq:14}) that the product of all $\inv$ eigenvalues (over all occupied bands at every symmetric momenta) has the same parity as $C_1$.

\begin{figure}
\centering
\includegraphics[width=8.5cm]{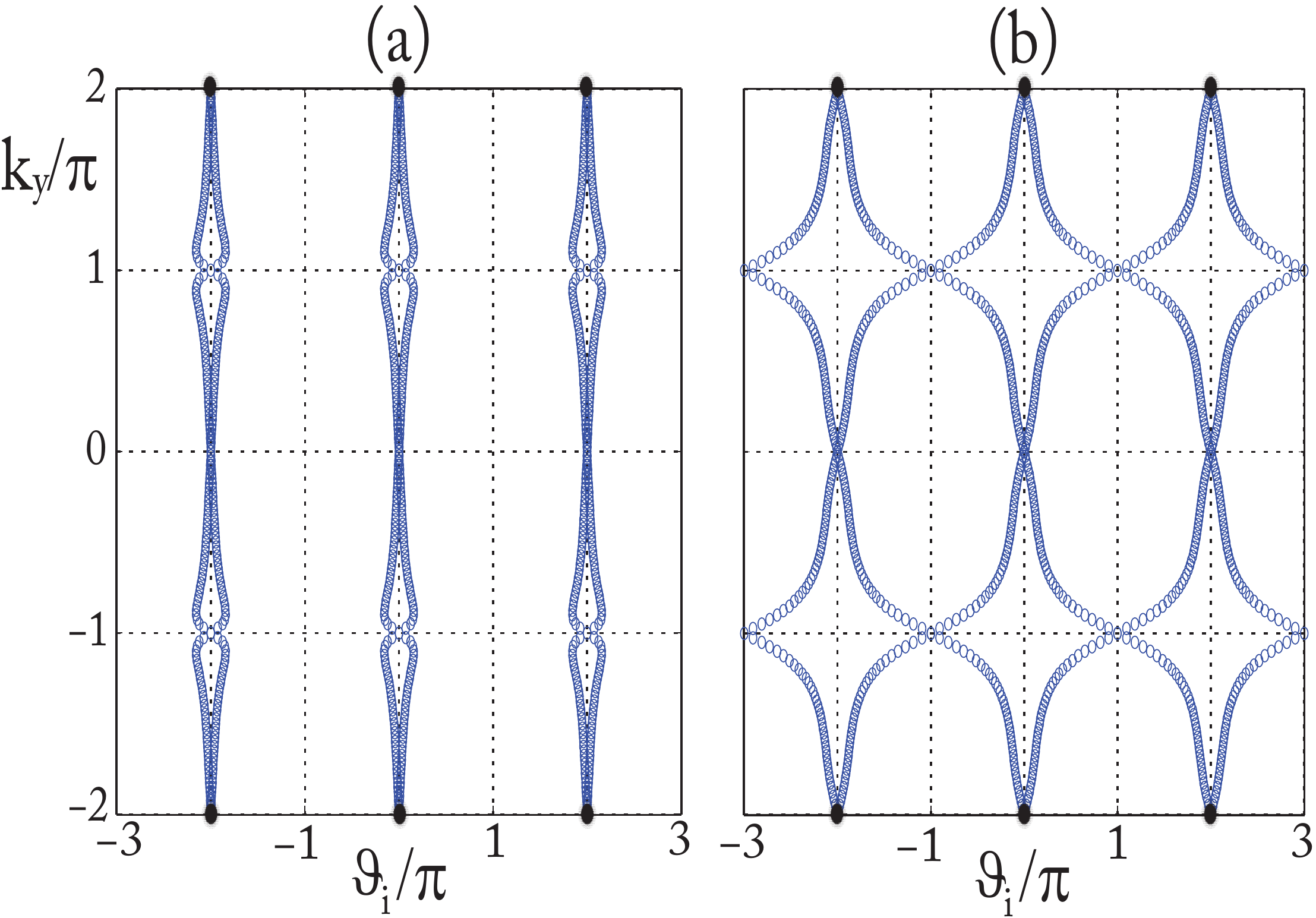}
\caption{The Wannier flow of a 2D insulator with both $\inv$ and TR symmetries. (a) This phase has a trivial TR-invariant; it is realized in the model Hamiltonian (\ref{eq:modeltrs}), with parameters $m=4.3$, $\delta=0$. (b) This phase has a nontrivial TR-invariant; it is realized in the same model, with $m=3$, $\delta=0$. }\label{fig:trstri}
\end{figure}

\subsection{The Inversion- and Time-Reversal-Symmetric Wilson Loop} \label{sec:invtrs}

Our aim is to highlight distinctive features of Wilson loops with both $\inv$ and time-reversal symmetries; we illustrate these features with a four-band model Hamiltonian:
\bal \label{eq:modeltrs}
h(k) \eq  (2-m-\text{cos}\, k_x - \text{cos}\,k_y)\,\Gamma_{03} + \delta \si k_y \;\Gamma_{12}   \lin
&+ \si k_x \;(\Gamma_{31}+\Gamma_{11}) +\si k_y \;( \Gamma_{21} + \Gamma_{02} )  ,
\end{align}
with matrices $\Gamma_{ij}$ defined as $\sigma_i \otimes \tau_j$; $\sigma_0$ ($\tau_0$) is the identity in spin (orbital) space; $\sigma_{i=1,2,3}$ ($\tau_{i=1,2,3}$) are Pauli matrices in spin (orbital) space. The Hamiltonian is $\inv$-symmetric: $\Gamma_{03} \,h(k)\,\Gamma_{03} = h(\text{-}k)$.   The Fermi energy is chosen so that there are two occupied bands in the ground state.  We tabulate the $\inv$ and $\W$-eigenvalues for different choices of parameters $m$ and $\delta$ in Tab. \ref{tab:trsmodel}. In this Section we set $\delta=0$, so the Hamiltonian is also time-reversal symmetric (TRS): $T \, h(k)\, T^{\text{-}1} = h(\text{-}k)$, with $T = i \Gamma_{20} \,K$; $K$ is the complex-conjugation operator. The two classes of TRS insulators are distinguished by a $\Z_2$ invariant $\Xi$, which is the change in time-reversal polarization over half an adiabatic cycle; $\Xi$ is odd for the nontrivial class.\cite{fu2006,kane2005A,kane2005B,bernevig2006a,bernevig2006c,koenig2007,qi2008B,wu2006,xu2006,fu2007b,moore2007,murakami2004A,roy2009, roy2009a,ran2008,hsieh2008,Prodan:2009oh,Prodan:2009mi,essin2009,jx2011,fu2007a,soluyanov2011} In Fig. (\ref{fig:trstri}-a) and (\ref{fig:trstri}-b), we have plotted the $\Wky$-spectra for both $\Z_2$-trivial $(m=4.3)$  and nontrivial $(m=3)$ phases respectively. \\

\begin{table}[h]
	\centering
		\begin{tabular} {|c|c|c|c|c|c|c|c|} \hline
		\multicolumn{2}{|c|}{Parameter} & \multicolumn{4}{|c|}{$\inv$ eigenvalues} & \multicolumn{2}{|c|}{$\W_{\kyi}$ eigenvalues}   \\ \hline
		 $\;m\;$ & $\;\delta\;$ &	$(0,0)$  & $ (\pi,0)$  & $ (0,\pi)$  & $ (\pi,\pi)$ & $\;\;\kyi=0\;\; $ & $ \kyi=\pi $  \\  \hline \hline
		$3$ & $0$ &  $(+ +)$ & $(+ +)$ & $(+ +)$ & $(- -)$ & $[+ +]$ & $[- -]$  \\ \hline
		 $3$ & $1$ &  $(+ +)$ & $(+ +)$ & $(+ +)$ & $(- -)$ & $[+ +]$ & $[- -]$  \\ \hline
		 $3$ & $2$ &  $(+ +)$ & $(+ +)$ & $(+ +)$ & $(- -)$ & $[+ +]$ & $[- -]$  \\ \hline
		 $4.3$ & $0$ &	$(+ +)$ & $(+ +)$ & $(+ +)$ & $(+ +)$ & $[+ +]$ & $[+ +]$ \\ \hline
		\end{tabular}
		\caption{For various choices of the parameters $m, \delta$ in the Hamiltonian (\ref{eq:modeltrs}), we write the corresponding (a) $\inv$ eigenvalues at the four symmetric momenta and (b) the eigenvalues of $\W_{\kyi}$. \label{tab:trsmodel}}
\end{table}

As derived in App. \ref{app:trs}, TRS imposes the following constraints on the $\Wky$ spectra: 

\noindent (i) The sets of eigenvalues at $\pm k_y$ are equal, \emph{i.e.},
\bal \label{eq:equalityset2dtrs}
\big\{ \text{exp}\,{i{\vartheta}_{k_y}}  \big\} = \big\{ \text{exp}\,i{\vartheta}_{\text{-}k_y} \big\}. 
\end{align}

\noindent (ii) The $\W$'s at symmetric momenta satisfy
\bal \label{eq:kramer}
\W_{\kyi}^{\text{-}1} = \Theta^{\text{-}1} \; \W_{\kyi} \; \Theta,
\end{align}
with $\Theta$ an antiunitary operator that squares to $\text{-}I$. This implies that every eigenstate of $\W_{\kyi}$ has a degenerate Kramer's partner. (i) and (ii) imply that if one Wannier center produces a Hall current $I_H$, its Kramer's partner produces a time-reversed current that cancels $I_H$.  As shown in Ref. \onlinecite{yu2011} and \onlinecite{soluyanov2011}, the $\Z_2$ invariant may be extracted from Fig. (\ref{fig:trstri}) in the following manner: in the region $k_y \in [0,\pi]$, $\vartheta \in [\text{-}\pi,\pi]$, let us draw a constant-$\vartheta$ reference line at any value of $\vartheta$. If the Wannier trajectories intersect this reference line an odd number of times, the phase is nontrivial, and vice versa.\\

By imposing $\inv$ symmetry as well, we arrive at the following conclusions:

\noindent (a) Due to $\inv$ symmetry, the $\W_{\kyi}$-spectra at $K_y=\{0,\pi\}$ consist of $\pm 1$ and complex-conjugate pairs; the additional constraint of Kramer's degeneracy implies that the spectra is composed of \emph{pairs} of $[\pone,\pone]$, \emph{pairs} of $[\text{-}1,\text{-}1]$ and complex-conjugate \emph{quartets} $[\lambda\,\lambda\,\lambda^*\lambda^*]$. Since time-reversal and $\inv$ commute, the two states in a Kramer's doublet must transform in the same representation under $\inv$ -- this limits the possible $\inv$ eigenvalues in a TRS ground state. 

\noindent (b) From (\ref{eq:equalityset2d}) and (\ref{eq:equalityset2dtrs}) we derive 
\bal 
\big\{ \text{exp}\,{i{\vartheta}_{k_y}}  \big\} = \big\{ \text{exp}\,{\text{-}i{\vartheta}_{\text{-}k_y}} \big\} = \big\{ \text{exp}\,{i{\vartheta}_{\text{-}k_y}} \big\},
\end{align}
which indicates that the flow of the Wannier centers in one quadrant, say $\vartheta \in [0,\pi]$ and $k_y \in [0,\pi]$, determines the flow in the full range, $\vartheta \in [\text{-}\pi,\pi]$ and $k_y \in [\text{-}\pi,\pi]$, by reflections. This is illustrated in Fig. (\ref{fig:trstri}), where each quadrant is bounded by dotted lines.

\begin{figure}
\centering
\includegraphics[width=8.5 cm]{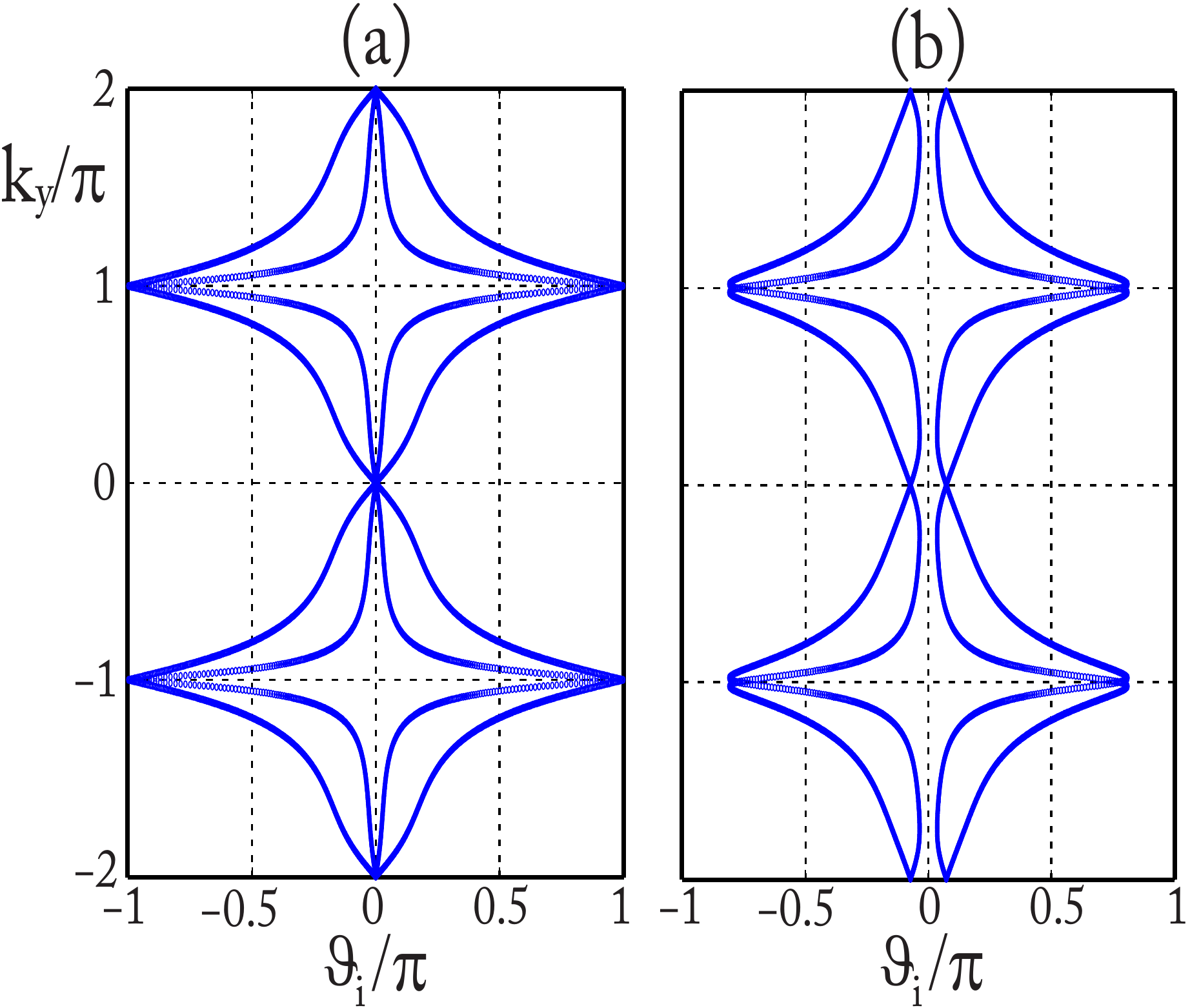}
\caption{ (a) An $\inv$-symmetric phase with relative winding $W=2$; spectral flow is protected by $\inv$ symmetry alone. This is the ground state of Hamiltonian (\ref{eq:modeltrs4band}). (b) Spectral flow is interrupted with an $\inv$-breaking perturbation. }\label{fig:W2C0Xi0vs}
\end{figure}

\subsection{Inversion-Protected Spectral Flow and the Relative Winding Number} \label{sec:invspec}

The Chern insulator and the TRS topological insulator exhibit spectral flow in the $\W$-spectrum; the symmetries that protect spectral flow are, respectively, charge conservation and time-reversal symmetry. In this Section, we report spectral flow of a third kind, which is protected by $\inv$ symmetry alone. In our first example, we consider the eight-band model
\bal \label{eq:modeltrs4band}
&h(k) =  (\text{-}1{-}\text{cos}\, k_x {-} \text{cos}\,k_y)\,\Theta_{030} + \si k_x \,(\Theta_{310}+\Theta_{110})   \lin
& +\text{sin}\, k_y \,( \Theta_{210} + \Theta_{020} ) +  0.8\, \text{sin}\, k_x \,(\Theta_{311}+\Theta_{111}),
\end{align}
with matrices $\Theta_{ijk}$ defined as $\sigma_i \otimes \tau_j \otimes \gamma_k$; $\sigma_{i=1,2,3}$ are Pauli matrices in spin space; for $i,j=\{1,2,3\}$, $\tau_{i}\otimes \gamma_{j}$ are products of Pauli matrices in a four-dimensional orbital space; $\sigma_0$ ($\tau_0 \otimes \gamma_0$) is the identity in spin (orbital) space. This Hamiltonian is $\inv$-symmetric: $\Theta_{030} \,h(k)\,\Theta_{030} = h(\text{-}k)$, and  time-reversal symmetric: $T \, h(k)\, T^{\text{-}1} = h(\text{-}k)$; here $T = i \Theta_{200} \,K$, and $K$ implements complex conjugation. The Fermi energy is chosen so that there are four occupied bands in the ground state. We tabulate the $\inv$ and $\W$-eigenvalues in Tab. \ref{tab:4bandW2}, and also plot the $\Wky$-spectrum in Fig. \ref{fig:W2C0Xi0vs}-a. With TRS, $C_1=0$. The change in time-reversal polarization over half an adiabatic cycle is $2$ (even), hence the TR invariant is trivial.\cite{fu2006} Yet, Wannier trajectories interpolate across the full unit cell: $\vartheta \in [0,2\pi)$. Let us softly break $\inv$ symmetry, while maintaining TRS, with the perturbation: $0.4\co k_x \,\Theta_{022} + 0.4 \co k_y \,\Theta_{112}$.  As evidence that spectral flow persists only with $\inv$ symmetry, we find in Fig. \ref{fig:W2C0Xi0vs}-b that the spectrum is now gapped.\\

\begin{table}[h]
	\centering
		\begin{tabular} {|c|c|c|c|c|c|} \hline
		 \multicolumn{4}{|c|}{$\inv$ eigenvalues} & \multicolumn{2}{|c|}{$\W_{\kyi}$ eigenvalues}   \\ \hline
		 	$(0,0)$  & $ (\pi,0)$  & $ (0,\pi)$  & $ (\pi,\pi)$ & $\;\;\kyi=0\;\; $ & $ \kyi=\pi $  \\  \hline \hline
		  $++++$ & $++++$ & $++++$ & $----$ & $++++$ & $----$  \\ \hline
				\end{tabular}
		\caption{For the Hamiltonian (\ref{eq:modeltrs4band}), we write the (a) $\inv$ eigenvalues at the four symmetric momenta and (b) the eigenvalues of $\W_{\kyi}$. \label{tab:4bandW2}}
\end{table}

\begin{figure}
\centering
\includegraphics[width=8 cm]{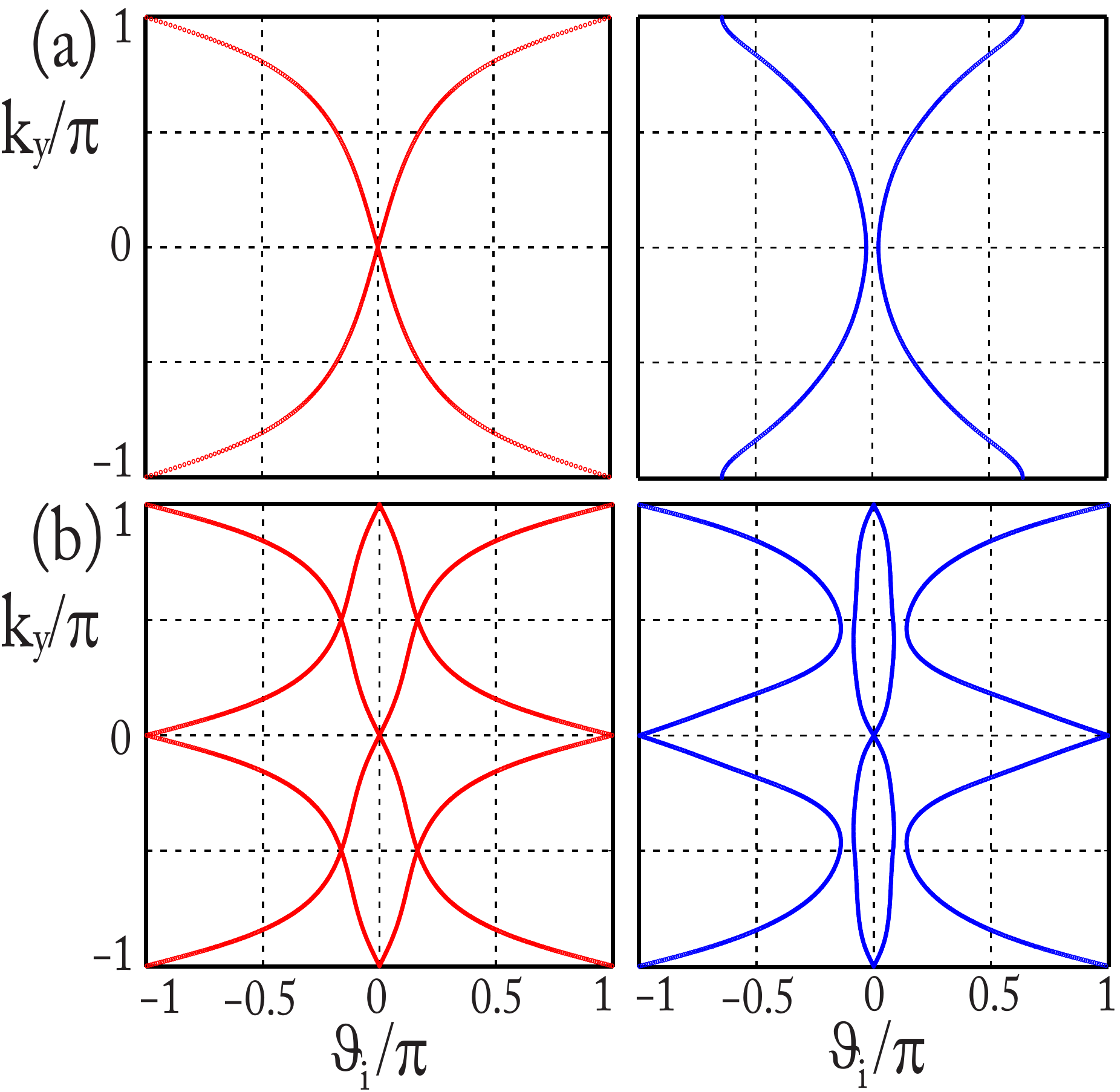}
\caption{ Figures in red provide two examples of accidental degeneracies. Upon a gap- and symmetry-preserving perturbation, these accidental degeneracies split, as shown in blue. These Wannier trajectories are calculated from 2D models; the Hamiltonians are not written explicitly. The model that describes the top two figures is a two-band model with $\inv$-eigenvalues equal to $\{+1,-1\}$ at all four inversion-invariant momenta. These are mapped to complex-conjugate $\W$-eigenvalues $[\lambda,\lambda^*]$, along both $k_y=0$ and $k_y=\pi$; \emph{cf.} Sec. \ref{sec:theorem1d}. $\lambda$ may vary in the interval $[1,-1]$, by continuous reparametrization of the Hamiltonian that maintains both energy gap and symmetry. }\label{fig:accidentaldegeneracies}
\end{figure}

$\inv$-protected spectral flow is characterized by a nonzero relative winding number $W$, which is defined in the following way. Two Wannier trajectories are said to wind relative to each other if they (i) intersect the \emph{same} primary site at a symmetric time $K_y$, then (ii) separate and intersect \emph{adjacent} secondary sites half a period later ($K_y+\pi$). $W$ is defined as the number of stable pairs of relatively-winding trajectories. Our definition relies only on $\inv$ symmetry, and does not depend on the presence or absence of any other symmetry. We outline a procedure to identify $W$:

\noi{a}  Count the number of Wannier trajectories that \emph{directly} connect mid-bond and primary sites in the quadrant $\{ \vartheta \in [0,\pi], k_y\in [0,\pi] \}$; call this number $n_1$. By a direct connection, we mean a smooth trajectory that flows without interruption. We consider three examples: in Fig. \ref{fig:accidentaldegeneracies}-a (blue) and \ref{fig:accidentaldegeneracies}-b (blue), $n_1=0$; $n_1=2$  in Fig. \ref{fig:W2C0Xi0vs}-a. 
 
\noi{b} Count the number of trajectories that directly connect mid-bond and primary sites in another quadrant  $\{ \vartheta \in [-\pi,0), k_y\in [0,\pi] \}$; call this $n_2$. In all three examples, $n_1=n_2$. 

\noi{c} $W$ is the minimum of $\{n_1,n_2\}$. We find that the insulator of Fig. \ref{fig:W2C0Xi0vs}-a has relative winding $W=2$; $W=0$ in the other two cases. 


\noindent For each  pair of relatively-winding trajectories, one of the pair $(\,\vartheta_1(k_y)\,)$ has winding number $\pone$ on the torus $\{ \vartheta \in [\text{-}\pi,\pi), k_y\in [\text{-}\pi,\pi) \}$, and the other has winding $\mo$. In principle, it is possible that $\vartheta_1(k_y)$ has winding $2n\pone$ ($n\in \Z^{\sma{+}}$), while its partner $\vartheta_2(k_y)$ has winding $\text{-}2n \mo$. However, $n>0$ implies that the trajectories $\vartheta_1(k_y)$ and $\vartheta_2(k_y)$ cross at a non-symmetric momentum ($k_y \neq \{0,\pi\}$); such degeneracies are not protected by symmetry - by a ground-state deformation that preserves both  the energy gap and $\inv$ symmetry, we may turn crossings into anti-crossings and thus reduce $n$ to $0$. Since $W$ is the number of \emph{stable} relatively-winding pairs, we eliminate all such accidental degeneracies \emph{before} carrying out the above procedure to identify $W$. All accidental degeneracies fall into two categories: (i) at non-symmetric momenta, there may be accidental crossings of two or more trajectories, as illustrated in Fig. \ref{fig:accidentaldegeneracies}-b (red). A slight deformation results in level repulsion, and turns crossings into anticrossings (Fig. \ref{fig:accidentaldegeneracies}-b (blue)). (ii) At symmetric momenta $\{0,\pi\}$, we rule out complex-conjugate-pair eigenvalues that are degenerate at either the primary or secondary site. In the example of Fig. \ref{fig:accidentaldegeneracies}-a (red), there is one such degeneracy at the primary site when $K_y=0$, and another at the secondary site when $K_y=\pi$; upon perturbing the Hamiltonian, this degeneracy splits, as shown in Fig. \ref{fig:accidentaldegeneracies}-a (blue).\\

\begin{figure}
\centering
\includegraphics[width=8.5 cm]{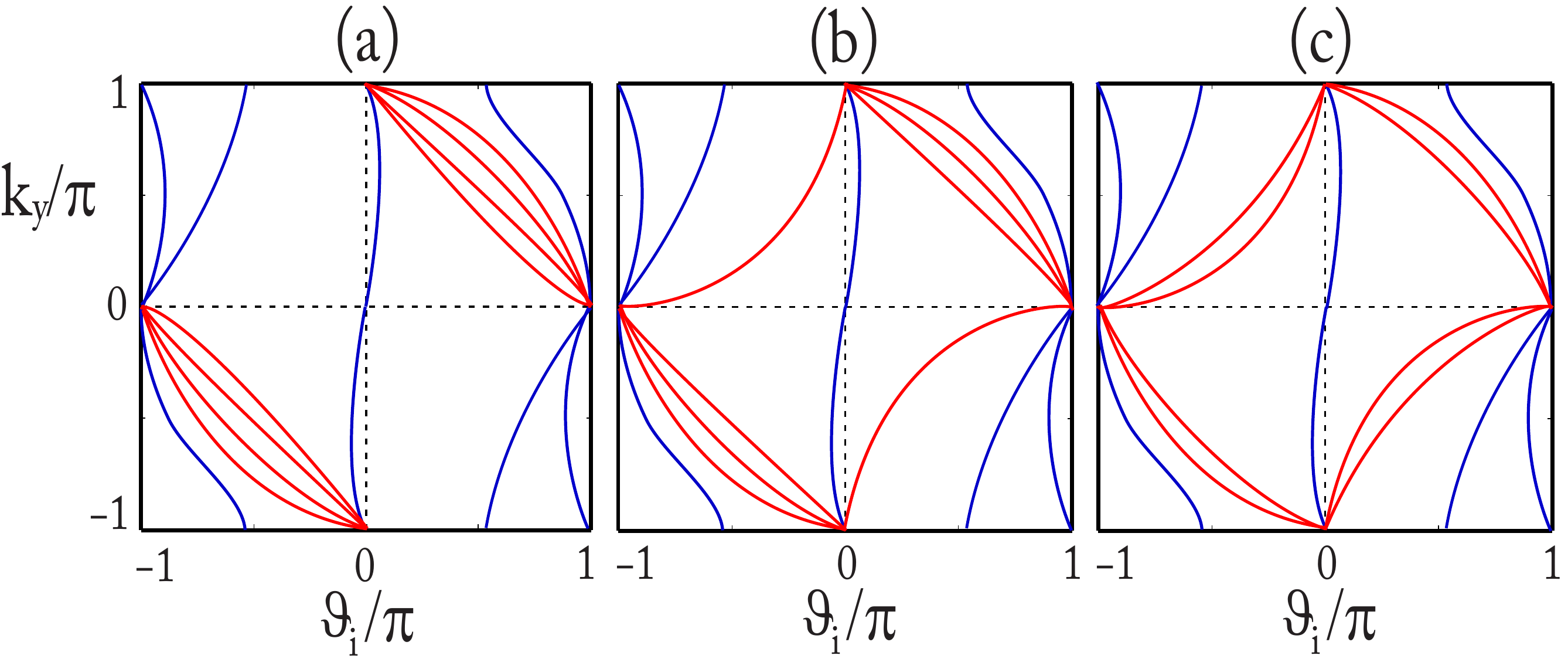}
\caption{Schematic illustrations of Wannier trajectories for eight occupied bands, where $N_d=4$.  (a) All four red trajectories wind with center-of-mass motion, thus $C_1=-4$, $W=0$. (b) One pair of red trajectories relatively wind, so $C_1=-2$, $W=1$. (c) Both pairs of red trajectories relatively wind, hence $C_1=0$, $W=2$.   }\label{fig:relativewinding}
\end{figure}

Let us derive a sufficient condition for relative winding, and simultaneously relate $W$, $C_1$ and the eigenvalues of $\W$ at constant $K_y=\{0,\pi\}$. We define $\Np(K_y)$, $\Nm(K_y)$ and $\Ncc(K_y)$ as the number of $\pone, \mo$ and complex-conjugate-pair eigenvalues of $\WKy$, respectively; $\Np(K_y)+\Nm(K_y)+\Ncc(K_y) = \noc$, the number of occupied bands. These quantities constrain the possible Wannier trajectories that interpolate between $K_y=0$ and $\pi$, and by implication they also constrain the possible topological invariants: $W$, $C_1$ and $\Xi$. Define $N_d$ as the maximum of two quantities:
\bal \label{eq:defineNd}
N_d = \text{max}\,&\big\{\;\Nm(\pi)-\Nm(0)-\Ncc(0),\lin
&\,\Nm(0)-\Nm(\pi)-\Ncc(\pi)\;\big\}.
\end{align}
If $N_d>0$, there are $N_d$ trajectories that directly connect the secondary site (at $K_y$) and the primary site (at $K_y+\pi$). Of these $N_d$ trajectories, one or more pairs may relatively wind, and the rest wind with center-of-mass motion, thus contributing to the Chern number $C_1$. The parity of $C_1$ is constrained as in (\ref{eq:w2flow}). A sufficient condition for relative winding is that $|C_1|<N_d$, in which case $2W = N_d - |C_1|$. If $|C_1| \geq N_d$, then $W=0$. In Fig. \ref{fig:relativewinding}, we schematically illustrate three cases where $N_d = \Nm(0)-\Nm(\pi)-\Ncc(\pi) = 7-1-2 = 4$, \emph{i.e.}, there are four trajectories (in red) that directly connect the secondary site (at $\pi$) and the primary site (at $0$). There are five ways to split four trajectories into center-of-mass and relative windings: $N_d = |C_1| + 2W =  |-4|+2(0) = |-2|+2(1) = |0| + 2(2) = |2|+2(1) = |4| + 2(0)$.  We illustrate the first three cases in Fig. \ref{fig:relativewinding}-a, b, and c respectively. For $\inv$-  and time-reversal-symmetric (I+TRS) insulators, $C_1=0$, hence the relative winding is related to $N_d$ through: $2W=N_d$. These relations, as summarized in Tab. \ref{tablegraphtheory}, also imply that $W$ is isotropic, in the following sense. We define $\{\varphi(k_x)\}$ as Wannier trajectories of the Wilson loop at constant $k_x$; if $\{\vartheta(k_y)\}$ exhibits relative winding $W$, then so will $\{\varphi(k_x)\}$. This claim is substantiated in  App. \ref{app:isotropy}.

\begin{table}[h]
	\centering
		\begin{tabular} {c|l|l|}  \cline{2-3}
			 &   $\;\;\;\;\;\;\;\;\;\;\;\;\;\;\;\; U(1)$  & $\;\;\;\;\;\;\;\;\;\;\;\;\;\;\;\;$ TRS  \\  \cline{2-3}  \hline 
		\multicolumn{1}{|c|}{$W>0$}  & $2W + |C_1| = N_d $ & $ W = \tfrac{1}{2}\,N_d$; \\ 
		 \multicolumn{1}{|c|}{} & & $\Xi = W$ mod $2$ \\ \hline
		\multicolumn{1}{|c|}{	} & $|C_1| \geq N_d$; & $\Xi = \tfrac{1}{2}\,\big(\Nm(0)-\Nm(\pi)\big)$  \\ 
		\multicolumn{1}{|c|}{$W=0$}	& $C_1 = \Nm(0)-\Nm(\pi)$  & $\;\;\;\;\;\;$ mod $2$ \\ 
		\multicolumn{1}{|c|}{}	& $\;\;\;\;\;\;\;\;$ mod $2$  &  \\ \hline
			\end{tabular}
		\caption{ Relations between relative winding $W \in \Zp$, the Chern number $C_1 \in \Z$, the TR invariant $\Xi \in \Z_2$, and eigenvalues of the Wilson loop at symmetric momenta. Columns: $U(1)$ denotes a generic insulator with charge-conservation symmetry; TRS denotes a time-reversal symmetric insulator. $N_d$ is defined in Eq. (\ref{eq:defineNd}). \label{tablegraphtheory}}
\end{table}



\subsubsection{Relative Winding of Insulators with both Inversion and Time-Reversal Symmetries}

In this Section we study the relative winding of insulators with both $\inv$ and TRS (I+TRS); we shall relate the relative winding $W$ with the TR invariant $\Xi$. While both $W$ and $\Xi$ characterize Wannier trajectories with no center-of-mass motion, they differ in many important respects. For I+TRS insulators with \emph{{nonzero}} relative winding, the parity of $W$ determines the TR invariant: $\Xi = W$ mod $2$; $\Xi$ is odd in the nontrivial class. To prove this, we apply the rule: modulo $2$, $\Xi$ equals the number of Wannier trajectories that intersect a constant-$\vartheta$ reference line.\cite{yu2011} With I+TRS, only one quadrant, e.g. $\{\vartheta \in [0,\pi], k_y \in [0,\pi]\}$, is independent. In the rest of this section, we denote coordinates in this quadrant by $(\vartheta,k_y)$. Two cases are possible: (i) $W$ number of trajectories directly connect points $(\vartheta,k_y)=(\pi,0)$ and $(0,\pi)$, or (ii) $W$ trajectories connect $(\pi,\pi)$ and $(0,0)$. If $\noc=2W$, there are exactly $W$ intersections with the reference line, hence $\Xi=W$ mod $2$. If $\noc > 2W$, it is possible in case (i) that:    (i-a) an extra trajectory connects $(0,0)$ and $(0,\pi)$, (i-b) a trajectory connects $(\pi,0)$ and $(\pi,\pi)$, (i-c) if there exists a complex-conjugate quartet $[\lambda_1 \lambda_1 \lambda_1^* \lambda_1^*]$ in the spectrum of $\W_{K_y=0}$, a pair of trajectories may connect $(0,\pi)$ with the complex-conjugate site at $(\lambda_1,0)$, and (i-d) if there exists a complex-conjugate quartet $[\lambda_2 \lambda_2 \lambda_2^* \lambda_2^*]$ in the spectrum of $\W_{K_y=\pi}$, a pair  of trajectories may connect $(\pi,0)$ with the complex-conjugate site $(\lambda_2,\pi)$. In all scenarios, these extra trajectories intersect an even number of times with the reference line -- the parity of the number of intersections is decided by $W$ alone. The proof is complete.\\ 

While only the parity of $W$ matters to the $\Z_2$ classification under TRS, $W$ provides a $\Zp$ classification under $\inv$ symmetry, and hence a more complete characterizaton. This distinction may be understood from a stability analysis of the Wannier centers. Since Kramer's degeneracy is two-fold, four or more Wannier centers generically experience level repulsion. Consider for example the $W=2$, I+TRS model of Fig. \ref{fig:W2C0Xi0vs}-a. Sitting at the primary site (at $K_y=\pi$) are four Wannier centers which are constrained by $\inv$-symmetry -- they do not experience level repulsion. If we now break $\inv$-symmetry while preserving TRS, these four Wannier centers destabilize and split to form two pairs of Kramer's doublets, thus breaking spectral flow; see Fig. \ref{fig:W2C0Xi0vs}-b.\\

\begin{figure}
\centering
\includegraphics[width=8 cm]{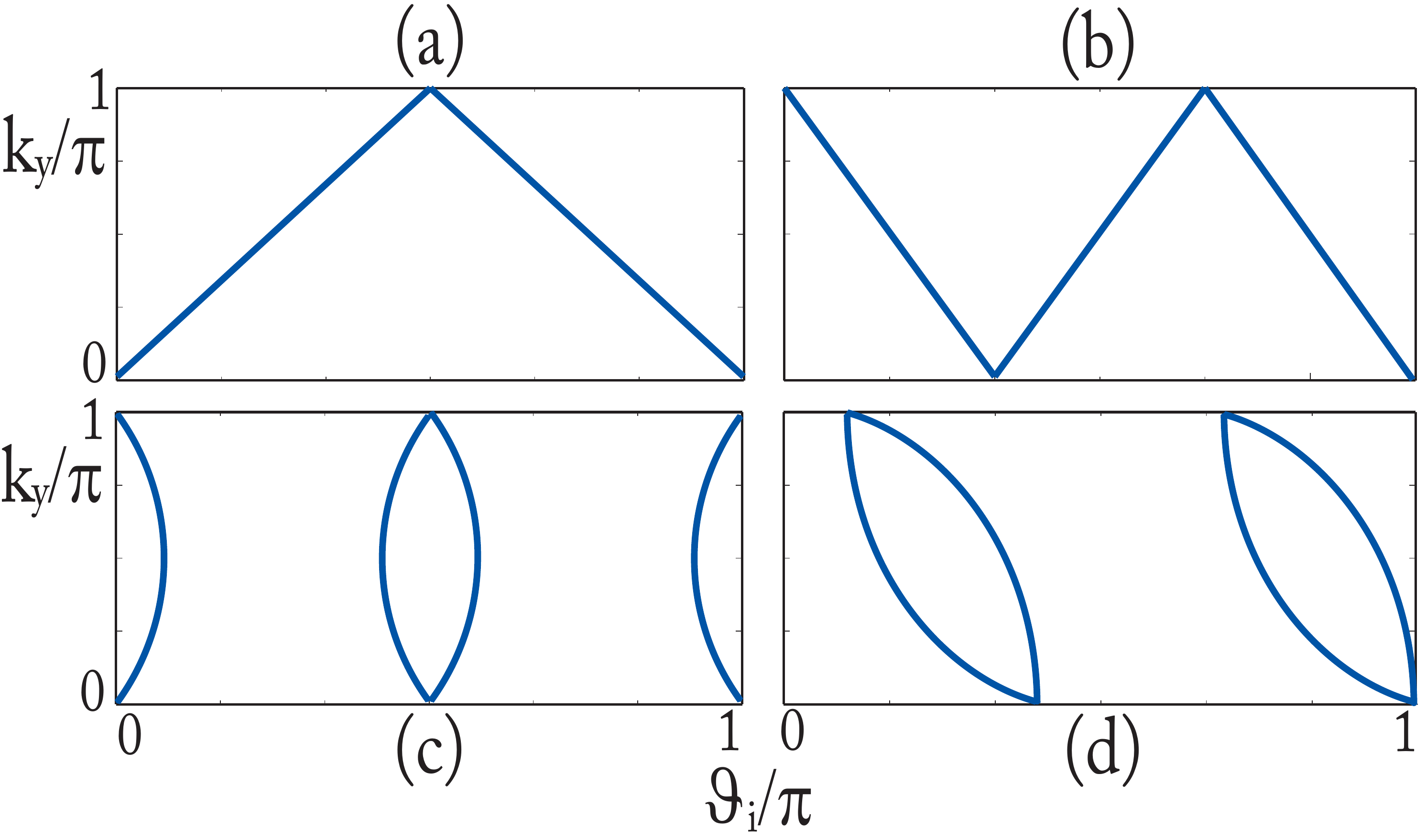}
\caption{ Schematic illustrations of wannier flows with zero relative winding. In figures (a) and (b), $\half (\Nm(0) -\Nm(\pi))=1$ (odd); the Wannier trajectories follow a zig-zag pattern. (c) $\half (\Nm(0) -\Nm(\pi))=0$ (even), and the Wannier flows are gapped. (d) $\half (\Nm(0) -\Nm(\pi))=2$ (even).   }\label{fig:zigzag}
\end{figure}

For I+TRS insulators with zero relative winding, it is possible that spectral flow is completely absent and the insulator is trivial. However, an I+TRS insulator with four or more occupied bands may have a nontrivial TR invariant without relative winding. To distinguish these two cases, we look to the indices $\Nm(0)$ and $\Nm(\pi)$.  Since $W=0$, all $\Nm(K_y)$ Wannier centers that originate from the secondary site (at $K_y$) must flow to either a complex-conjugate site or a secondary site (at $K_y + \pi$). By conservation of trajectories, $\Nm(K_y)-\Nm(K_y+\pi)$ number of trajectories must connect the secondary site (at $K_y$) to complex-conjugate sites (at $K_y+\pi$); here we have assumed $\Nm(K_y)>\Nm(K_y+\pi)$. Since $C_1=0$, we know from (\ref{eq:w2flow}) that $\Nm(K_y)-\Nm(K_y+\pi)$ is even -- within one quadrant, e.g., $\{\vartheta \in [0,\pi]$, $k_y \in [K_y,K_y+\pi]\}$, $(\Nm(K_y)-\Nm(K_y+\pi))/2$ number of trajectories connect the secondary site (at $K_y$) to  complex-conjugate sites (at $K_y+\pi$). Two cases arise: (i) if $(\Nm(K_y)-\Nm(K_y+\pi))/2$ is \emph{odd}, there must be at least \emph{one other} trajectory that interpolates between $K_y$ and $K_y+\pi$. Then the sum of all trajectories that connect to complex-conjugate sites is \emph{even}, as required by Kramer's degeneracy. Hence, TRS enforces a zig-zag pattern of Wannier flows, as illustrated schematically in Fig \ref{fig:zigzag}-a and -b. (ii) For even $(\Nm(K_y)-\Nm(K_y+\pi))/2$, Kramer's degeneracy is satisfied without additional trajectories, and the resultant Wannier flows are gapped, as in Fig. \ref{fig:zigzag}-c and -d. Therefore, 
\bal \label{eq:RconstrainsXi}
\Xi  \eq \tfrac{1}{2}\,\big(\,\Nm(0) -\Nm(\pi)\,\big) \;\;\text{mod}\;\;2 \lin
\eq \half \,N_d \;\;\text{mod}\;\;2;
\end{align}
for the last equality, we applied the definition of $N_d$ in Eq. (\ref{eq:defineNd}) and that $\Ncc(K_y)$ is a multiple of four; \emph{c.f.} Sec. \ref{sec:invtrs}. The relation (\ref{eq:RconstrainsXi}) applies to I+TRS insulators with $W>0$ as well, since we have proven $\Xi= W$ mod $2$ in this Section, and previously identified $2W = N_d$ in Sec. \ref{sec:invspec}. The relations between $\Xi$, $W$ and $N_d$ are summarized in Tab. \ref{tablegraphtheory}.\\

\begin{figure}
\centering
\includegraphics[width=8.5 cm]{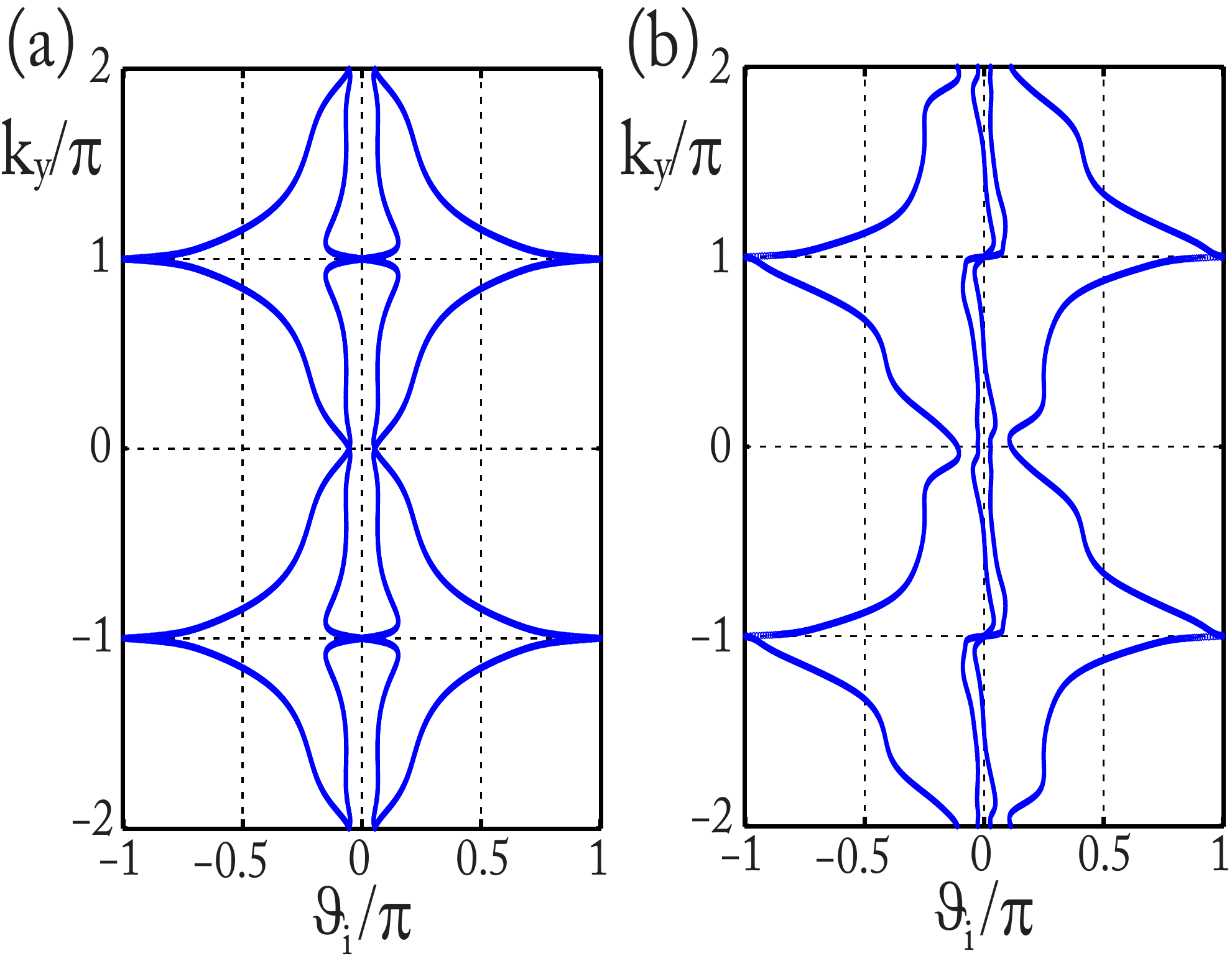}
\caption{ (a) An $\inv$-symmetric phase with zero relative winding; spectral flow persists due to TRS alone. This is the ground state of Hamiltonian (\ref{eq:modelW0}). (b) Spectral flow is interrupted with a TRS-breaking perturbation. }\label{fig:C0W0_nontrivialXivstrivial}
\end{figure}

Consider for example the eight-band model:
\bal \label{eq:modelW0}
&h(k) = (\,-2-\co k_x -\co k_y\,)\,\Theta_{033} + \tfrac{4}{5}\,\co k_y\,\Theta_{001} \lin
&+5 \si k_x \,\big( \Theta_{113} + \Theta_{313} \big)    -   \tfrac{3}{2}\si k_y\,\big(\Theta_{020} +\Theta_{210}\big) \lin
& + (\,6\si k_x +\tfrac{7}{2}\,\si k_y\,)\,\big(\,\Theta_{023} + \Theta_{213}\,\big) + \Theta_{030},
\end{align}
with matrices $\Theta_{ijk}$ defined in (\ref{eq:modeltrs4band}). This Hamiltonian has the same symmetries as that in (\ref{eq:modeltrs4band}), namely $\inv$ and TRS, and the ground state is defined to be the four lowest-energy bands. We tabulate the $\inv$ and $\W$-eigenvalues in Tab. \ref{tab:4bandW0XiTop}. As illustrated in Fig. \ref{fig:C0W0_nontrivialXivstrivial}-a, the relative winding is trivial, but partner-switching occurs with help from the complex-conjugate quartet at $K_y=0$. In this $W=0$ model, spectral flow is protected by TRS alone. As evidence, we deform the Hamiltonian with a TRS-breaking term, $\Delta h(k) = 8\si k_x\,\big(\,\Theta_{022} + \Theta_{120} + \Theta_{121}\,\big) + 2\si k_y \,\Theta_{123}$,
which preserves both the energy gap and $\inv$ symmetry. As illustrated in Fig. \ref{fig:C0W0_nontrivialXivstrivial}-b, Kramer's degeneracy is now lifted -- the quartet $[\lambda \lambda \lambda^* \lambda^*]$ at $K_y=0$ splits into two separate doublets $[\lambda_1\, \lambda_1^*], [\lambda_2\, \lambda_2^*]$. This contrasts with a previous example, where we broke TRS in a $\Xi$-nontrivial phase; as shown in Fig. \ref{fig:windingnumberfigures}-b, the resultant $\W$-spectrum is not gapped, due to a nonzero relative winding.

\begin{table}[h]
	\centering
		\begin{tabular} {|c|c|c|c|c|c|} \hline
		 \multicolumn{4}{|c|}{$\inv$ eigenvalues} & \multicolumn{2}{|c|}{$\W_{\kyi}$ eigenvalues}   \\ \hline
		 	$(0,0)$  & $ (\pi,0)$  & $ (0,\pi)$  & $ (\pi,\pi)$ & $\;\;\kyi=0\;\; $ & $ \kyi=\pi $  \\  \hline \hline
		  $++--$ & $++--$ & $++--$ & $----$ & $\lambda\,\lambda\,\lambda^*\lambda^*$ & $++--$  \\ \hline
				\end{tabular}
		\caption{For the Hamiltonian (\ref{eq:modelW0}), we write the (a) $\inv$ eigenvalues at the four symmetric momenta and (b) the eigenvalues of $\W_{\kyi}$. \label{tab:4bandW0XiTop}}
\end{table}

\section{Discussion} \label{app:conclusion}

We have formulated a $\Zp$ classification of 1D and 2D insulators with inversion ($\inv$) symmetry. Since the Berry phase and inversion eigenvalues of a Bloch band depend on the choice of unit cell, the 1D classification is useful so long as there is a preferred unit cell. In monatomic Bravais lattices, the natural unit cell is centered around the atom. If there are multiple atoms per unit cell, a natural choice of the unit cell may not exist, unless such choice is selected by the presence of an edge. In this sense, the 1D Berry phase reflects boundary physics. An experimentally-relevant example is the boundary charge theorem of Vanderbilt, which relates the charge on an edge to the bulk polarization.\cite{vanderbilt1993} Yet another example lies in the {entanglement} spectrum of 1D insulators, where a spatial entanglement cut mimics an edge.\cite{li2008,peschel2003} \\

Our results have implications for experiments that directly measure Berry phase through interference. These experiments require coherent transport across the BZ, as has been realized in semiconductor superlattices \cite{wannierbook,mendez1993,resta2000,sundaram1999,chang2008,mendez1988} and optical lattices of cold atoms.\cite{raizen1997,niu1996, dahan1996, wilkinson1996} Due to the unit cell ambiguity in translationally-invariant systems, the Berry phases are determined only up to a global phase. In 1D, this implies that only differences in Berry phases are physical observables; such differences have been measured by Atala \emph{et. al.} in a cold-atom setting.\cite{atala2013} On the other hand, if a 2D insulator exhibits spectral flow in the Berry phases, such a property is invariant under a global translation of phases. Thus, the relative winding of an $\inv$-symmetric insulator is physically observable.\\

\emph{Acknowledgements}
AA is grateful to Taylor L. Hughes, Chen Fang, Yang-Le Wu, Alexey Soluyanov, Dmitry Abanin and  Edward Witten for illuminating discussions. AA was supported by the First Year Fellowship at Princeton University and NSF CAREER DMR-095242. XD was supported by NSF of China and the 973 program of China (No. 2007CB925000). BAB was supported by Princeton Startup Funds, NSF CAREER DMR-095242, ONR - N00014\text{-}11\text{-}1-0635 and  NSF-MRSEC DMR-0819860 at Princeton University. This work was also supported by DARPA under SPAWAR Grant No.: N66001\text{-}11\text{-}1-4110.

\appendix

\section{Derivation of parallel transport equation (\ref{eq:paralleltransport})} \label{app:paralleltransport}

Let us define $u^n_k$ as eigenfunctions of the flat-band Hamiltonian ${\cal H}_{\tex{F}}(k)$, with energy $\varepsilon_{\text{-}}$; the Hamiltonian is derived in Eq. (\ref{eq:flatband2}). $n$ is a band index that runs over the occupied bands, and $k$ labels the crystal momentum.  $u^n_k$ form basis vectors in the subspace of occupied bands. Upon adiabatic evolution through the Brillouin zone, the state $u^n_{k^{\sma{(i)}}}$ at some initial momentum $k^{\sma{(i)}}$ (at time $t_i$) is mapped to a different state $v^n_k$ at final momentum $k$ (at time $t$). In this process, the state acquires a dynamical phase $\int^t_{t_i} \varepsilon_{\text{-}}dt$. In addition, there is a unitary rotation of basis vectors in the occupied subspace; this rotation is affected by a $U(\noc)$  Wilson-line matrix $\W_{k \leftarrow k^{\sma{(i)}}}$: 
\bal
&v^n_k(r) = e^{-i\int^t_{t_i} \varepsilon_{\text{-}}dt} \sum_{m=1}^{\noc} \,u^m_k(r)\,\W^{mn}_{k \leftarrow k^{\sma{(i)}}}
\end{align}
Let us apply the time-dependent Schrodinger equation:
\bal
{\cal H}_{\tex{f}}(k)\,v^n_k(r) = i\partial_t \big[e^{-i\int^t_{t_i} \varepsilon_{\text{-}}dt}  \,\sum_m\,u^m_k(r)\,\W^{mn}_{k \leftarrow k^{\sma{(i)}}} \big].
\end{align}
In the adiabatic approximation, $v^n_k$ remains an eigenstate of the ${\cal H}_{\tex{F}}(k)$ with energy $\varepsilon_{\text{-}}$. Upon cancellation of the dynamical phase and replacing $\partial_t = \sum_{\mu=1}^d \,(dk_{\mu}/dt)\, \partial_{k_{\mu}}$ for a $d$-dimensional BZ, we arrive at Eq. (\ref{eq:paralleltransport}).

\section{The Tight-Binding Formalism and the Relation between the Full and Tight-Binding Connections} \label{app:connections}

In the tight-binding variational approximation, the Hilbert space is reduced to $\ntot$ atomic orbitals $\phi_{\alpha}(r-R-r^{(\alpha)})$; $\phi_{\alpha}$ are eigenstates of an atomic Hamiltonian, and $\alpha=1,2,\ldots,\ntot$ denotes orbital and spin. $R$ is a Bravais lattice vector that denotes a unit cell. Within each unit cell, $r^{(\alpha)}$ is the position of the atom corresponding to orbital $\alpha$. Due to the finite overlap of atomic orbitals, we are motivated to construct an orthonormal basis that preserves the point-group symmetries of $\phi_{\alpha}$. Such a basis is realized with \low functions,\cite{slater1954,goringe1997,lowdin1950} which are defined as
\bal \label{eq:lowdin}
\pdg{\varphi}_{\alpha}(x-R_i-r^{(\alpha)}) = \sum_{j,\beta} \pdg{\phi}_{\beta}(x-R_j-r^{(\beta)})\;[\Delta^{\text{-}1/2}]^{\beta \alpha}_{ji}
\end{align}
with the hermitian overlap integral
 \bal
 \Delta^{\ab}_{ij} = \int d^{\tex{d}}r\; \phi^*_{\alpha}(x-R_i-r^{(\alpha)})\;\pdg{\phi}_{\beta}(x-R_j-r^{(\beta)}).
 \end{align}
Here, $i,j$ are indices for Bravais lattice vectors; we sum over repeated indices. The orthogonality of \low function reads as
\bal \label{eq:lowortho}
\int d^dr \,\varphi^*_{\alpha}(x-R_i-r^{(\alpha)})\,\pdg{\varphi}_{\beta}(x-R_j-r^{(\beta)}) = \delta_{\ab}\,\delta_{i,j}.
\end{align} 
Let us define the inversion operator $\hat{I}$ by its action on functions: $\hat{I}\,f(r) = f(\text{-}r)$. With discrete translational symmetry in $d$ spatial dimensions, there exists $2^d$ inversion centers which are not related by lattice translations. Many results in this paper are derived for a unit cell that is inversion-symmetric. By this we mean that (i) the spatial origin coincides with one of the $2^d$ inversion centers, and (ii) the set of atoms within the unit cell are closed under the inversion operation, \emph{i.e.}, for each orbital $\phi_{\alpha}$ located at $r^{(\alpha)}$ in the unit cell, such orbital is mapped by inversion to one or more orbitals $\phi_{\beta}$ at $r_{\beta}$, such that $r_{\beta}= -r_{\alpha}$. We choose this unit cell for analytic convenience -- a different choice results in a global translation of Berry phases, as shown in App. \ref{app:wilsonprojection}. To formalize (ii), one may define a Hermitian, unitary, $\ntot \times \ntot$  sewing matrix $\wp$ in the basis of atomic orbitals:
\bal \label{eq:overlapmatrix}
\pdg{\wp}_{\ab} = \int d^dr\; {\phi}^*_{\alpha}(r-r^{(\alpha)}) \,\hat{I}\,\pdg{\phi}_{\beta}(r-r^{(\beta)}).
\end{align}
The \low functions $\varphi$, which are constructed from $\phi$ through (\ref{eq:lowdin}), transform identically under point-group symmetry operations,\cite{slater1954} hence $\hat{I}\,\pdg{\varphi}_{\alpha}(r-r^{(\alpha)}) =  \pdg{\varphi}_{\delta}(r-r^{(\delta)})\,\pdg{\wp}_{\delta \alpha}$ as well. In Hamiltonians with discrete translational symmetry, we form the Bloch sums
\bal
\pdg{u}_{k \alpha}(r) = \tfrac{1}{\sqrt{N}} \sum_R e^{-ik(r-R-r^{(\alpha)})} \varphi_{\alpha}(r-R-r^{(\alpha)}),
\end{align}
which are periodic in lattice translations. The tight-binding Hamiltonian (\emph{cf.} (\ref{eq:physH})) is defined as $[h(k)]_{\alpha \beta} = \big\langle \pdg{u}_{k \alpha} \big| \;{\cal H}(k) \;\big| \pdg{u}_{k\beta}\big\rangle$. In the tight-binding approximation, the periodic component of the Bloch wave $\psi^n_k$ is $u^n_k(r) = \sum_{\alpha}  \,[U^n_k]_{\alpha}\,\pdg{u}_{k \alpha}(r)$, where  $[U^n_k]_{\alpha}$ is the unitary eigenmatrix that diagonalizes $h(k)$. Up to a gauge transformation, the tight-binding Hamiltonian is periodic in reciprocal lattice vectors $G$:
\bal
h(k+G) = V(G)^{\mo}\,h(k)\,V(G),
\end{align}
where $V(G)$ is a unitary matrix with elements: $[V(G)]_{\ab} = \delta_{\ab}\,e^{iGr^{(\alpha)}}$. We define the periodic gauge as 
\bal \label{periodicgauge}
\ket{U^m_{k+G}} = V(G)^{\mo}\,\ket{U^m_k},
\end{align}
which is the tight-binding equivalent of the continuum gauge condition: $u^n_{k+G}(r) = u^n_k(r)\,e^{-iG\cdot r}$. A different choice of spatial origin, \emph{e.g.} $\forall \alpha, \;\; r_{\alpha} \rightarrow r_{\alpha}+\delta$,  results in a global phase change $V(G) \rightarrow V(G)\,e^{iG \cdot \delta}$.

\begin{widetext}

We define 
\bal \label{eq:positionop}
[\hat{X}_{\mu}]^{nm}_k = \frac{1}{{N}}\sum_{\ab}[U^{n}_k]_{\beta}^* [U^m_k]_{\alpha}\sum_{R,R'}e^{ik(R+r^{(\alpha)}-R'-r^{(\beta)})} \int d^{\tex{d}}r    \varphi^*_{\beta}(r-R'-r^{(\beta)})  [\,r_{\mu}-R_{\mu}-r_{\mu}^{(\alpha)}\,] \pdg{\varphi}_{\alpha} (r-R-r^{(\alpha)}), 
\end{align}
where $\mu$ denotes a spatial direction. By inserting the expression for $u^n_k$ into (\ref{eq:fullconn}), we show that the continuum connection $C_k$ (cf. (\ref{eq:fullconn})) differs from the tight-binding connection $A_k$ (cf. (\ref{eq:tightconn})): $C_{\mu}(k) = A_{\mu}(k) - i \,[\hat{X}_{\mu}]_k$. Thus, the eigenspectra of (\ref{eq:Wilsonloopdefined}) and (\ref{eq:wilsonfull}) generically differ.

\end{widetext}

\section{Properties of Wilson loops} \label{app:wilsonprojection}

In this Appendix we derive several properties of Wilson loops, which apply to both continuum and tight-binding versions, as defined in   Sec. \ref{sec:holonomy} and \ref{sec:tightWilson}  respectively. The derivations have been carried out with the tight-binding connection $A_k$ (\emph{cf.} Eq. (\ref{eq:tightconn}) ) and for a 1D BZ; they are trivially generalizable to the continuum connection $C_k$ (\emph{cf.} Eq. (\ref{eq:fullconn}) ) and to higher dimensions.

\noindent (i) Let us denote a Wilson line, over a path  with start point $k^{\sma{(i)}}$ and end point $k^{\sma{(f)}}$,  as $\W_{k^{\sma{(f)}}\leftarrow k^{\sma{(i)}}}$. The Wilson line between two infinitesimally-separated momenta is ${\cal W}^{mn}_{k + \epsilon \leftarrow k}   =   \bra{U^m_{k+\epsilon}} U^n_k \big\rangle$. Let us define  loop $\cal{L}$ as tracing a path between base point $k^{\sma{(0)}}$ and end point $k^{\sma{(0)}} + 2\pi$;  we divide the $\calL$ into infinitesimally-separated momenta: $\{k^{\sma{(0)}} + 2\pi,k^{\sma{(N)}},k^{\sma{(N\text{-}1)}} \ldots k^{\sma{(2)}}, k^{\sma{(1)}},k^{\sma{(0)}}\}$, with $N \gg 1$. Let ${\cal P}_{k} = \sum_{m=1}^{\noc} \,\ket{U^m_k}\bra{U^m_k}$ denote the projection to the occupied bands at momentum $k$. The Wilson loop may be discretized: $[{\cal W}(\calL)]^{mn} $
\bal \label{eq:finitetosmall}
\eq {\cal W}^{ml}_{k^{\sma{(0)}}+2\pi \leftarrow k^{\sma{(N)}}}\;{\cal W}^{lq}_{k^{\sma{(N)}}\leftarrow k^{\sma{(N\text{-}1)}}} \ldots \;{\cal W}^{op}_{k^{\sma{(2)}}\leftarrow k^{\sma{(1)}}}\;{\cal W}^{pn}_{k^{\sma{(1)}}\leftarrow k^{\sma{(0)}}}  \lin
\eq \bra{U^m_{k^{\sma{(0)}}+2\pi}}  \;{\displaystyle \prod_{\alpha}^{k^{\sma{(0)}}+2\pi \leftarrow k^{\sma{(0)}}}}\, {\cal P}_{k^{(\alpha)}} \;\ket{U^n_{k^{\sma{(0)}}}} .
\end{align}
The product of projections are path-ordered (symbolized by $\leftarrow$) along ${\cal L}$, with the earlier-time momenta positioned to the right.

\noindent (ii) Let ${\cal C}$ be a path that connects two arbitrary momenta $k^{\sma{(1)}}$ and $k^{\sma{(2)}}$, and ${\cal C}^{\sma{T}}$ be the same path with opposite path-ordering. The Wilson line satisfies the unitary condition
\bal \label{eq:unitaryline}
\dg{\W_{k^{\sma{(2)}}\leftarrow k^{\sma{(1)}}}({\cal C})} = \W_{k^{\sma{(1)}}\leftarrow k^{\sma{(2)}}}({\cal C}^{\sma{T}}) = \W_{k^{\sma{(2)}}\leftarrow k^{\sma{(1)}}}({\cal C})^{\text{-}1}.
\end{align}
For $k^{\sma{(2)}}=k^{\sma{(1)}}+2\pi$, the above relation generalizes to Wilson loops. From (\ref{eq:unitaryline}) one can derive that for a fixed loop, the eigenspectrum of $\W$ is independent of the base point.

\noindent (iv) We insist on the periodic gauge condition (\ref{periodicgauge}). It follows that the non-Abelian Berry factors are the unimodular eigenvalues of the operator
\bal \label{manifest}
\hat{\W} = V(G)\,{\displaystyle \prod_{q}^{k+2\pi \leftarrow k}}\, {\cal P}_{q}.
\end{align}
Since each projection is invariant under a $U(\noc)$ gauge transformation, the eigenspectrum of $\W$ is manifestly gauge-invariant.

\noindent(v) We noted in App. \ref{app:connections} that translating the spatial origin by $\delta$ results in $V(G) \rightarrow V(G)\,e^{iG \cdot \delta}$. From (\ref{manifest}) we deduce that the eigenspectrum of $\W$ is translated by a global $U(1)$ phase.

\section{Wilson Loops and the Projected Position Operator} \label{app:projectedposition}

Let us demonstrate that the phases of the $\W$-spectrum coincide with the eigenspectrum of the projected position operator ${\cal P}^{\tex{occ}}\,\hat{x}\,{\cal P}^{\tex{occ}}$.\cite{kivelson1982b} Here, ${\cal P}^{\tex{occ}}$ projects to the occupied subspace of the translationally-invariant Hamiltonian (\ref{eq:genhamil}), which have Bloch eigenfunctions $\psi^n_k(x) = e^{ikx}\,u^n_k(x)$. We are interested in eigenfunctions $\Psi$ that satisfy
\bal \label{eq:eigenposition}
\big(\,{\cal P}^{\tex{occ}}\,\hat{x}\,{\cal P}^{\tex{occ}} - \tfrac{1}{2\pi}\,\vartheta\,\big)\;\Psi(x) =0
\end{align} 
for some eigenvalue $\vartheta/2\pi$. We expand $\Psi$ in the subspace of occupied Bloch waves:
\bal
\Psi(x) = \sum_{n=1}^{\noc}\,\int \tfrac{dk}{2\pi} \;f_{nk}\;\psi_{k}^n(x).
\end{align} 
In the periodic gauge ($\psi^n_k = \psi^n_{k+2\pi}$), the action of the projected position operator on the wavefunction $f$ may be decomposed into an intra-band operator $\partial_k$ and an inter-band operator ${C}$:\cite{Blount}
\bal
\bra{\psi_{k}^n} \;{\cal P}^{\tex{occ}}\,\hat{x}\,{\cal P}^{\tex{occ}} \;\ket{\Psi} = i\, \frac{\partial f_{nk}}{\partial k} + i\,\sum_{m=1}^{\noc}\,{[C(k)]}^{nm}\,f_{mk},
\end{align}
where $C$ is the non-Abelian connection that is defined in (\ref{eq:fullconn}). In the general solution to (\ref{eq:eigenposition}), the wavefunctions at two different momenta ($k$ and $k^{\sma{(i)}}$) are related by the Wilson line
\bal
f_{mk} = e^{-i(k\text{-}k^{\sma{(i)}}) \vartheta/2\pi}\;\sum_{n=1}^{\noc}\;\big[\;\text{T} \;e^{ \text{-}\int^k_{k^{\sma{(i)}}} \, C(q) \,dq}\;\big]^{mn}\; f_{nk^{\sma{(i)}}}.
\end{align}
It follows from the periodic boundary condition on $f$ that
\bal \label{eq:eigenwilson}
\sum_{n=1}^{\noc}\;\W^{mn}_{k+2\pi \leftarrow k}\;f_{nk} &= e^{i\vartheta}\;f_{m,k+2\pi} = e^{i\vartheta}\;f_{m,k}, 
\end{align}
hence we have shown that the eigenspectrum of ${\cal P}^{\tex{occ}}\,\hat{x}\,{\cal P}^{\tex{occ}}$ coincides with the phases of the eigenspectrum of $\W$. Furthermore, let us derive the eigenfunctions of ${\cal P}^{\tex{occ}}\,\hat{x}\,{\cal P}^{\tex{occ}}$. Define the $U(\noc)$ matrix $Q(k)$ such that its columns are the eigenstates of the Wilson loop at base point $k$: $\W_{k+2\pi \leftarrow k}$. That is, $\W_{k+2\pi \leftarrow k} = Q(k)\,D\,\dg{Q}(k)$,
where $D$ is a diagonal matrix that contains the eigenvalues of $\W$. While the eigenvalues of $\W$ do not depend on the base point $k$, the eigenfunctions $Q(k)$ do. The matrix $Q(k)$ is related to the matrix $Q(k'\neq k)$ by a Wilson line:
\bal \label{eq:qkrelation}
&\W_{k'+2\pi \leftarrow k'} = \dg{\W}_{k \leftarrow k'}\;\W_{k+2\pi \leftarrow k} \;\W_{k \leftarrow k'} \lin
&\imp Q(k') = \dg{\W}_{k \leftarrow k'}\;Q(k) = \W_{k' \leftarrow k}\;Q(k).
\end{align}
We label the $j$'th diagonal element in $D$ by $\text{exp}\,(\,i\vartheta^{(j)}\,)$, where $j$ runs from $1,2 \,\ldots\,\noc$. If $\vartheta^{(j)}/2\pi$ is an eigenvalue of ${\cal P}^{\tex{occ}}\,\hat{x}\,{\cal P}^{\tex{occ}}$, so is any integer addition to $\vartheta^{(j)}/2\pi$, as is consistent with $\vartheta^{(j)}$ being a phase. For each occupied band, there exists an infinite ladder of eigenvalues: $\vartheta^{(j)}_R/2\pi = \vartheta^{(j)}/2\pi + R \,;  R \in \mathbb{Z}.$ The index $R$ labels the unit cell where the eigenfunction is localized. We will choose the convention that  $\text{-}\pi\,<\,\vartheta^{(j)}\,\leq\, \pi$. Let us then label the eigenfunctions of ${\cal P}^{\tex{occ}}\,\hat{x}\,{\cal P}^{\tex{occ}}$ by $j=1 \,\ldots\,\noc$ and $R \in \mathbb{Z}$. The wavefunction $[f^{\sma{(j)}}_{\sma{R}}]_{n,k} \propto Q(k)_{nj}$ so as to satisfy the eigenvalue equation (\ref{eq:eigenwilson}). In addition, we multiply the $j$'th column of the eigenmatrix $Q(k)$ by a momentum-dependent phase, so as to ensure periodicity:
\bal
&[f^{(j)}_{R}]_{n,k+2\pi} =e^{-i(k+2\pi)\vartheta^{(j)}_{R}/2\pi}\;Q(k+2\pi)_{nj} \lin
\eq e^{-i(k+2\pi)\vartheta^{(j)}_{R}/2\pi}\;\sum_{m=1}^{\noc} \W^{nm}_{k+2\pi \leftarrow k}\;Q(k)_{mj} \lin
\eq  e^{-i(k+2\pi)\vartheta^{(j)}_{R}/2\pi}\;e^{i\vartheta^{(j)}}\;Q(k)_{nj} = [f^{(j)}_{R}]_{n,k}.
\end{align}
In the second equality, we have applied the relation (\ref{eq:qkrelation}). In summary, to each eigenvalue $\vartheta^{(j)}_{R}/2\pi$ there corresponds an eigenfunction of ${\cal P}^{\tex{occ}}\,\hat{x}\,{\cal P}^{\tex{occ}}$:
\bal \label{eq:generalizedwann}
\Psi^{(j)}(x\text{-}R) = \sum_{n=1}^{\noc} \int \tfrac{dk}{2\pi} \;e^{\text{-}ik\,(\tfrac{\vartheta^{(j)}}{2\pi}+R)}\;Q(k)_{nj}\;\psi_{k}^n(x).
\end{align}

\begin{widetext}

\section{Analytic Properties of the Sewing Matrix and Constraints on the Wilson Loop due to Inversion Symmetry} \label{app:1dconstraint}

Let us employ the bra\text{-}ket notation that is introduced in Sec. \ref{sec:tightWilson}. Inversion symmetry constrains the Hamiltonian as $\wp \,h(k)\,\wp = h(\text{-}k)$, where $\wp$ is an overlap matrix that is defined in Eq. (\ref{eq:overlapmatrix}). This implies that the occupied eigenstates $U^m_k$ at $\pm k$ are related through inversion by a $U(\noc)$ `sewing matrix' $B_k$:
\bal \label{eq:sewU}
[U^m_{\text{-}k}]_{\alpha} = \sum_{n=1}^{\noc} \sum_{\beta} \;B^{mn*}_k\,\pdg{\wp}_{\alpha \beta}\,  [U^n_k]_{\beta}.
\end{align}
Here, $m,n$ are indices that label the occupied bands. Alternatively, $ B^{mn}_k = [U^m_{\text{-}k}]_{\alpha}^*\;\pdg{\wp}_{\alpha \beta}\;\pdg{[U^n_k]}_{\beta} = \bra{U^m_{\text{-}k}}\;\wp\;\ket{U^n_k}$. We define the inversion-invariant momenta $\ki$ as satisfying $-\ki = \ki + G(\ki)$ for some reciprocal lattice vector $G$. Implicit in the Wilson loop is the periodic gauge condition (\ref{periodicgauge}), from which we find $B^{mn}_{\ki}=\bra{U^m_{\ki}}\,V(G(\ki))\,\wp\,\ket{U^n_{\ki}}$. For an inversion-symmetric unit cell, as defined in (\ref{app:connections}), it follows that $\wp\,V(G)\,\wp^{\mo} = V(-G)$, and hence $(\,V(G(\ki))\,\wp\,)^2 = \wp^2 = I$, thus the inversion operator $V(G(\ki))\,\wp$ has eigenvalues $\pm 1$. It follows from  $\wp \,h(k)\,\wp = h(\text{-}k)$ that $[\,h(\ki),V(G(\ki))\,\wp\,]=0$, thus it is possible to find a basis in which $U^n_{\ki}$ are  eigenstates of $V(G(\ki))\,\wp$. Then $B_{\ki}$ is a diagonal matrix with diagonal elements equal to the $\inv$ eigenvalues of $U^n_{\ki}$. In deriving the gauge-invariant $\W$-eigenspectrum, we will employ such a convenient gauge. We prove that $B_k$ is unitary. Let us label the unoccupied bands by the primed index $m'=\noc \pone, \ldots, n_{\tex{tot}}$. If the ground state is insulating, \emph{i.e.}, $\bra{U^{m'}_{\text{-}k}}\;\wp\;\ket{U^n_k}=0$ for all $n$ occupied bands and all $m'$ unoccupied bands, 
\bal
&\sum_{m=1}^{\noc}\,B^{mn}_k\,B^{ml*}_k = \sum_{m=1}^{\noc} \bra{U^m_{\text{-}k}}\,\wp\,\ket{U^n_k}\,\bra{U^m_{\text{-}k}}\,\wp\,\ket{U^l_k}^* = \sum_{m=1}^{n_{\tex{tot}}} \bra{U^l_{k}}\,\wp\,\ket{U^m_{\text{-}k}}\,\bra{U^m_{\text{-}k}}\,\wp\,\ket{U^n_k} = \bra{U^l_{k}}\,\wp^2\,\ket{U^n_k} = \delta^{ln}.
\end{align}
In the second equality, we have used the identity: $\bra{U^{m'}_{\text{-}k}}\;\wp\;\ket{U^n_k}=0$; the third equality required the completeness relation. Furthermore, the Hermicity of the inversion operation $\wp$ implies $B_{\text{-}k}= \dg{[B_k]} = B_k^{\text{-}1}$. Applying (\ref{eq:sewU}) and $\wp^2=I$, we derive the following condition for a Wilson line between two infinitesimally-separated momenta $k^{\sma{(1)}}$ and $k^{\sma{(2)}}$:
\bal
\W_{\text{-}k^{\sma{(2)}} \leftarrow \text{-}k^{\sma{(1)}}}^{mn} = \braket{U^m_{\text{-}k^{\sma{(2)}}}}{U^n_{\text{-}k^{\sma{(1)}}}}  =B^{mo}_{k^{\sma{(2)}}}\, \bra{U^o_{k^{\sma{(2)}}}}\,\wp^2\,\ket{U^l_{k^{\sma{(1)}}}} \,B^{nl*}_{k^{\sma{(1)}}} = B^{mo}_{k^{\sma{(2)}}}\,[\WL]_{k^{\sma{(2)}} \leftarrow k^{\sma{(1)}}}^{ol} \,B^{nl*}_{k^{\sma{(1)}}}.
\end{align} 
This relation is generalizable to a Wilson line between arbitrary momenta. We divide the finite path between $k^{\sma{(1)}}$ and $k^{\sma{(2)}}$ into infinitesimally-separated momenta: $\{k^{\sma{(1)}}, k^{\sma{(1)}}+\Delta, k^{\sma{(1)}}+2\Delta,\ldots,k^{\sma{(2)}}-\Delta,k^{\sma{(2)}}\}$. Following Eq. (\ref{eq:finitetosmall}),
 \bal \label{eq:Wilsonsewing}
&\W_{k^{\sma{(2)}} \leftarrow k^{\sma{(1)}}}^{mn} = \W_{k^{\sma{(2)}} \leftarrow k^{\sma{(2)}}-\Delta}^{ml} \,\W_{k^{\sma{(2)}}-\Delta \leftarrow k^{\sma{(2)}}-2\Delta}^{lo}\,\ldots\,\W_{k^{\sma{(1)}}+\Delta \leftarrow k^{\sma{(1)}}}^{pn} \lin
& =B^{mr}_{\text{-}k^{\sma{(2)}}}\,\W_{\text{-}k^{\sma{(2)}} \leftarrow \text{-}k^{\sma{(2)}}+\Delta}^{rs} \,B^{ls*}_{\text{-}k^{\sma{(2)}}+\Delta}\,B^{lt}_{\text{-}k^{\sma{(2)}}+\Delta}\,\W_{\text{-}k^{\sma{(2)}}+\Delta \leftarrow \text{-}k^{\sma{(2)}}+2\Delta}^{tu} \,B^{ou*}_{\text{-}k^{\sma{(2)}}+2\Delta}\, \ldots\, B^{pv}_{-\Delta}\,\W_{\text{-}k^{\sma{(1)}}-\Delta \leftarrow \text{-}k^{\sma{(1)}}}^{vw} \,B^{nw*}_{\text{-}k^{\sma{(1)}}}\lin
& =B^{mr}_{\text{-}k^{\sma{(2)}}}\,\W_{\text{-}k^{\sma{(2)}} \leftarrow \text{-}k^{\sma{(2)}}+\Delta}^{rs} \,\W_{\text{-}k^{\sma{(2)}}+\Delta \leftarrow \text{-}k^{\sma{(2)}}+2\Delta}^{so}  \ldots \W_{\text{-}k^{\sma{(1)}}-\Delta \leftarrow \text{-}k^{\sma{(1)}}}^{pw} \,B^{nw*}_{\text{-}k^{\sma{(1)}}} = B^{mr}_{\text{-}k^{\sma{(2)}}}\,\W_{\text{-}k^{\sma{(2)}} \leftarrow \text{-}k^{\sma{(1)}}}^{rs}\,B^{ns*}_{\text{-}k^{\sma{(1)}}}. 
\end{align}
Applying this result to a Wilson loop between $\text{-}\pi$ and $\pi$,
\bal \label{eqq}
\W_{\pi \leftarrow \text{-}\pi}^{mn} = B^{mr}_{\text{-}\pi}\,{{\W}^{ rs}_{ \text{-}\pi \leftarrow \pi}}\,B^{ns*}_{\pi} = B^{mr}_{\text{-}\pi}\,{{\W}^{\dagger rs}_{ \pi \leftarrow \text{-}\pi}}\,B^{ns*}_{\pi} = B^{mr}_{\text{-}\pi}\,{{\W}^{\dagger rs}_{ \pi \leftarrow \text{-}\pi}}\,\dg{[B^{sn}_{\text{-}\pi}]}.
\end{align}
Here we have used the identities (\ref{eq:unitaryline}) and  $B_{\ki} = \dg{B_{\ki}}$. (\ref{eqq}) informs us that $\W$  is equivalent to its Hermitian adjoint through a unitary transformation, \emph{i.e.}, the \emph{set} of $\W$-eigenvalues is equal to its complex conjugate:
\bal \label{eq:setequality1d}
\big\{ \text{exp}\,{i\vartheta}  \big\} = \big\{ \text{exp}\,{\text{-}i\vartheta} \big\}. 
\end{align}
Let us derive a second useful identity. The Wilson loop along $\calL$ may be decomposed into two Wilson lines that each connect two symmetric momenta: $\W(\calL) = \W_{\;{\pi \leftarrow 0}} \,\W_{\;{0 \leftarrow \text{-}\pi}}$. Up to a change in orientation, $\W_{\;{\pi \leftarrow 0}}$ is mapped to $\W_{\;{0 \leftarrow \text{-}\pi}}$ by an inversion $k \rightarrow \text{-}k$. Applying (\ref{eq:Wilsonsewing}),
\bal \label{eq:Wilsonloop1D}
\W_{\pi \leftarrow \text{-}\pi}^{mn} = \W_{\pi \leftarrow 0}^{ml}\,\W_{0 \leftarrow \text{-}\pi}^{ln} = B^{mr}_{\pi}\,[{\W}^{\dagger}_{ 0\leftarrow \text{-}\pi}]^{rs}\,B^{sl}_{0}\,\W_{0 \leftarrow \text{-}\pi}^{ln}.
\end{align}
We have derived (\ref{eq:setequality1d}) and (\ref{eq:Wilsonloop1D}) for the tight-binding Wilson loop ( \emph{cf.} (\ref{eq:Wilsonloopdefined}) ), for which the connection involves tight-binding eigenfunctions $U^n_k$. Similar results apply to the continuum Wilson loop (  \emph{cf.} (\ref{eq:wilsonfull}) ), for which the connection involves $u^n_k$ --  eigenfunctions of the Bloch Hamiltonian (\ref{eq:blochhamil}). We may similarly define a $U(\noc)$ sewing matrix: $\bar{B}^{mn}_k = \int d^dr \;u^{m*}_{\text{-}k}(r)\,u^n_k(\text{-}r)$ that links Bloch eigenfunctions at $\pm k$. Within the tight-binding approximation, we may relate $u^n_k$ and $U^n_k$ through App. \ref{app:connections}, and identify  $\bar{B}$ as identical to the sewing matrix $B$ of (\ref{eq:sewU}). This identity implies that the mapping between $\inv$ and $\W$ eigenvalues, as presented in Sec. \ref{sec:theorem1d}, applies to both tight-binding and continuum Wilson loops. If the ground state is characterized by $n_s=0$ ($n_s$ is defined in Sec. \ref{sec:theorem1d}), the spectra of both Wilson loops are identical, and comprise only $\pm 1$ eigenvalues. If $n_s>0$, the spectra of both Wilson loops comprise the same numbers of $\pm 1$ and complex-conjugate-pair eigenvalues; they may differ only in the phases of the complex eigenvalues.

\end{widetext}

\section{Case Studies of the 1D Inversion-Symmetric Insulator} \label{app:casestudy}

\subsection{Case Study: $1$ Occupied Band}
 
With one occupied band, there is only one Wannier center per unit cell. In each unit cell, the Wannier center can only be at a primary site or at a secondary site; only these spatial configurations in a periodic lattice are invariant under inversion. If the Bloch waves at $k=0$ and $\pi$ transform under different representations of $\inv$, $\W=\text{-}1$. Proof: We shall employ notation that is defined in (\ref{eq:Wilsonloop1D}). Being unitary, $\W_{\;{0 \leftarrow \text{-}\pi}}$ must be of the form $\text{exp}\,({i\vartheta})$. (\ref{eq:Wilsonloop1D}) informs us that  $\W({\calL}) = B_{\pi}\,\text{exp}\,({\text{-}i\vartheta})\,B_0\,\text{exp}\,({i\vartheta}) = B_{\pi}\,B_0$, which is a product of the $\inv$ eigenvalues of the sole occupied band at $k=0$ and $\pi$. The proof is complete. This result can be verified by a model tight-binding Hamiltonian
\bal
h(k) = {-}(\alpha + \co k )\; \tau_3 + \si k \;\tau_2,
\end{align}
with $\tau_i$ defined as Pauli matrices in orbital space. This Hamiltonian has an $\inv$ symmetry: $\tau_3 \,h(k)\, \tau_3 = h(\text{-}k)$; the insulator is trivial when $\alpha>1$, and nontrivial when $\text{-}1<\alpha<1$.

\subsection{Case Study: $2$ Occupied Bands} \label{sec:casestudy2band}

For $\inv$-symmetric insulators with two occupied bands, $\W$ is a $2 \times 2$ matrix; our case study captures many qualitative features of larger-dimensional $\W$'s. We assume a general form for $\W_{\;{0 \leftarrow \text{-}\pi}}$ that satisfies unitarity:
\bal \label{eq:unitaryform}
\W_{\;{0 \leftarrow \text{-}\pi}} \eq e^{i\alpha}\;\begin{pmatrix} c &d \\ \text{-}d^* & c^* \end{pmatrix}\;;\;\;\; |c|^2 + |d|^2 =1.
\end{align}
Inserting (\ref{eq:unitaryform}) into (\ref{eq:Wilsonloop1D}), we arrive at $\W(\calL) =$
\bal \label{eq:2DWilsonparametrization}
 \begin{pmatrix} \xi^1_{\pi}\big(|c|^2\xi^1_0 + |d|^2\xi^2_0\big) && c^*d\,\xi^1_{\pi}\big(\xi^1_0 - \xi^2_0\big) \\ cd^*\,\xi^2_{\pi}\big(\xi^1_0 - \xi^2_{0}\big)& & \xi^2_{\pi}\big(|c|^2\xi^2_0 + |d|^2\xi^1_0\big) \end{pmatrix},
\end{align}
where $\xi^1_{\ki}$ and $\xi^2_{\ki}$ are diagonal elements of the sewing matrix $B_{\ki}$. We exhaust the possible $\inv$ eigenvalues $\{ \xi_{\ki}\}$, solve the characteristic equations and derive the $\W$ spectra. Our results are tabulated in Tab. \ref{table1d2band}. 
  
Cases (i)-(iii) of Tab. \ref{table1d2band} may be summarized as: if the $\inv$ eigenvalues of occupied bands at either $k=0$ or $\pi$ are identical, \emph{i.e.}, if either sewing matrix $B_0$ or $B_{\pi}$ is proportional to the identity, then the $\W$-spectrum comprises only $\pm 1$ eigenvalues; its eigenvalues are the diagonal elements of the product $B_0\,B_{\pi}$.

In case (iv) of Tab. \ref{table1d2band}, we encounter (a) occupied bands with nonidentical $\inv$ eigenvalues at both $k=0$ and $\pi$,  {and} (b) a complex-conjugate pair of $\W$-eigenvalues ($\lambda \lambda^*$): a pair of Wannier centers are positioned equidistantly on opposite sides of each primary site.  The exact position is not determined by symmetry; this arbitrariness reflects the range in this equivalence class of $\inv$-symmetric insulators, \emph{i.e.}, it is possible to tune the Hamiltonian and sweep the interval of allowed $\lambda$ while preserving both the insulating gap and $\inv$ symmetry. This implies that case (iv) is trivial. Why? Let us tune the Hamiltonian to a limit in which bands of identical representation are fully coupled, \emph{i.e.}, the positive-$\inv$ (negative-$\inv$) band at $k=\text{-}\pi$ adiabatically evolves into the positive-$\inv$ (negative-$\inv$) at $0$, through the Wilson line $\W_{\;{0 \leftarrow \text{-}\pi}}$. For example, if $\xi^1_0=\xi^2_{\pi} = \text{-}\xi^1_{\pi} =\text{-}\xi^2_{0}$, the two $\W$-eigenvalues are $|d|^2-|c|^2 \pm 2i|c||d|$. Tuning $|c|\rightarrow 0$ and $|d|\rightarrow 1$ effectively decouples the two-band $\W$ into two one-band $\W$'s; each Abelian $\W$ connects bands of the same representation between $k=0$ and $\pi$, hence each contributes $\pone$ to the spectrum, and $\W \rightarrow I$. As discussed in the Introduction, if $\W$ is tunable to the identity, then the insulator is in the same equivalence class as the atomic insulator, hence we conclude that case (iv) is trivial. 

In contrast with complex-conjugate-pair $\W$-eigenvalues, protected $\text{-}1$ eigenvalues obstruct $\W$ from being tuned to the identity. The nontrivial insulators are cases (ii) and (iii) of Tab. \ref{table1d2band}, which have one and two $\text{-}1$ eigenvalues respectively. We distinguish between case (iii) and the degenerate limit of case (iv), in which $\W \rightarrow \text{-}I$; this is the limit in which bands of identical representation are fully decoupled, \emph{i.e.}, the positive-$\inv$ (negative-$\inv$) band at $k=\text{-}\pi$ adiabatically evolves into the negative-$\inv$ (positive-$\inv$) at $0$. In case (iii), the equality $\W=\text{-}I$ is robust against gap- and symmetry-preserving transformations of the ground-state, while this is not true in case (iv).

\begin{table}[t]
	\centering
		\begin{tabular} {|c|c|c|c|c|} \hline
			\multicolumn{2}{|c|}{Model parameters} & \multicolumn{2}{|c|}{$\inv$ eigenvalues}    & $\W$ \\ \cline{1-4} 
			$\;\;\;\;\;\;\alpha \;\;\;\;\;\;$ & $\beta$ & $0$ & $\pi$ & spectrum \\ \hline \hline
			$0$ & $0$ & $\;(+ +)\;$&$(+ +)$ & $[+ +]$  \\ \hline
			$1.5$ & $0$ & $(+ -)$&$(+ +)$ & $[+ -]$ \\ \hline
			$1.5$ & $\text{-}1.5$ & $(- -)$&$(+ +)$ & $[- -]$ \\ \hline
			$1.5$ & $1.5$ & $(+ -)$&$(+ -)$ & $[\lambda \lambda^*]$ \\ \hline
		\end{tabular}
		\caption{A model of an $\inv$-symmetric ground state with two occupied bands: for various choices of the parameters $\alpha,\beta$ in the Hamiltonian (\ref{eq:model2occ}), we write the corresponding $\inv$ and $\W$-eigenvalues. \label{tab:model2occ}}
\end{table}

Let us verify our results in Tab. \ref{table1d2band} with the model, tight-binding Hamiltonian
\bal \label{eq:model2occ}
& h(k) =  -\Gamma_{03} + 0.1\,\Gamma_{13} + (\Gamma_{21}+\Gamma_{31}) \si k \lin
& \; +\half \big(\, \alpha \,(\Gamma_{30}+\Gamma_{03}) + \beta \, (\Gamma_{30}-\Gamma_{03}) \,\big) \co k,
\end{align}
with matrices $\Gamma_{ij}$ defined as $\sigma_i \otimes \tau_j$; $\sigma_0$ ($\tau_0$) is the identity in spin (orbital) space; $\sigma_{i=1,2,3}$ ($\tau_{i=1,2,3}$) are Pauli matrices in spin (orbital) space. The Hamiltonian is $\inv$-symmetric: $\Gamma_{03} \,h(k)\,\Gamma_{03} = h(\text{-}k)$.   The Fermi energy is chosen so that there are two occupied bands in the ground state. We tabulate the $\W$ spectra for various choices of the parameters $\alpha$ and $\beta$ in Tab. \ref{tab:model2occ}.

\section{Proof of mapping between $\inv$ and $\W$ eigenvalues} \label{app:1d4occbands}

We employ notation that has been defined in Sec. \ref{sec:theorem1d}: $n_{\sma{ (\pm)}}(\ki)$, FBOP, $k_s, \xi_s, n_s$. As defined in Eq. (\ref{eq:sewU}), $B_k$ is a $U(\noc)$ sewing matrix linking bands at $\pm k$ through $\inv$. At symmetric momenta $\ki$, a gauge is chosen in which $B_{\ki}$ is a  diagonal matrix with elements $\{\xi_1(\ki),\xi_2(\ki), \ldots, \xi_{\noc}(\ki)\}$ equal to the $\inv$ eigenvalues at $\ki$. At momentum $k_s$, we pick a convention that the fewest bands of one parity (FBOP) are indexed by $m=1,2,\ldots,n_s$ and the rest of the bands at $k_s$ are indexed by $m=n_s\pone,\ldots,\noc$. We evaluate the Wilson loop $\W$ over a path that begins at $k_s\text{-}\pi$, sweeps the interval $[k_s\text{-}\pi, k_s + \pi]$ and ends at $k_s+\pi$; the $\W$-spectrum is independent of the base point, as shown in App. \ref{app:wilsonprojection}. In the rest of the section, we simplify notation and assume that $\W = \W_{k_s+\pi \leftarrow k_s\text{-}\pi}$. Let us reproduce a result presented in   (\ref{eq:Wilsonloop1D}): $\W = B_{k_s+\pi} \,\dg{Z}\, B_{k_s}\, Z$,
with $Z$ defined as the Wilson line $\W_{k_s \leftarrow k_s\text{-}\pi}$; $Z^{-1}=\dg{Z}$. Let us define $Y$ as the $\noc \times \noc$ matrix
\bal \label{eq:Pdefine}
Y = \tfrac{1}{2} \big(I + \xi_s\, B_{\pi+k_s}\, \W \big).
\end{align}
Since $B_{\pi+k_s}^2 = I$, the matrix elements of $Y$ are
\bal \label{eq:Pelements}
Y_{ij} = \tfrac{1}{2} \big(I + \xi_s \,\dg{Z}\, B_{k_s}\, Z\big)_{ij} = \sum_{l=1}^{n_s}\, Z_{li}^*\;Z_{lj}.
\end{align}
Here we have applied the unitarity condition $\dg{Z}{Z}=I$  to express $[\dg{Z} B_{k_s} Z]_{ij} = \sum_{a=1}^{\noc} \; \xi_a(k_s)\; Z_{ai}^*\;Z_{aj} = \text{-}\xi_s(\delta_{ij} - 2\,\sum_{b=1}^{n_s} Z_{bi}^*\;Z_{bj})$. We deduce from (\ref{eq:Pelements}) that $Y$ is a rank-$n_s$ projection matrix. If $n_s=0$, $Y$ is the zero matrix. We define $\bar{Y}^{\alpha_1 \alpha_2 \ldots \alpha_m}$ as $m \times m$ submatrices in $Y$ that lie on the intersections of rows numbered by $\{\alpha_1, \alpha_2, \ldots, \alpha_m\}$ with columns numbered by $\{\alpha_1, \alpha_2, \ldots, \alpha_m\}$. For example, $\bar{Y}^{1} = \sum_{a=1}^{n_s} Z_{a,1}^*\;Z_{a,1}$, and
\bal
\bar{Y}^{23} = \begin{pmatrix}  \sum_{a=1}^{n_s} Z_{a,2}^*\;Z_{a,2} & & \sum_{a=1}^{n_s} Z_{a,2}^*\;Z_{a,3} \\ \sum_{a=1}^{n_s} Z_{a,3}^*\;Z_{a,2} & & \sum_{a=1}^{n_s} Z_{a,3}^*\;Z_{a,3} \end{pmatrix}. 
\end{align}
The determinant of a $m \times m$ submatrix $\bar{Y}$ is also called the $m \times m$ minor of $Y$. According to a well\text{-}known theorem in linear algebra, the rank of a matrix is equal to the largest integer $r$ for which a nonzero $r \times r$ minor exists, therefore $\text{det}\;\bar{Y}^{\alpha_1 \alpha_2 \ldots \alpha_{m}} =0\;\text{if} \;m > n_s.$ Applying (\ref{eq:Pdefine}), the characteristic equation $\text{det}[\lambda\;I - \W ] = 0$ is equivalent to 
\bal \label{eq:characequiv}
0 \eq \text{det} [\;\text{-}\xi_s \,B_{k_s +\pi}\;] \;\text{det}[\;\lambda\;I - \W \;] \lin
\eq \text{det} [ \;( \text{-}\xi_s\,\lambda \,B_{k_s +\pi} - I) +2\,Y \;]
\end{align}

\begin{widetext}
We claim that the determinant in the second line of (\ref{eq:characequiv}) is equal to a polynomial in $\lambda$ of order $2n_s$, multiplied by the factor  $(\text{-}\xi_s\, \lambda \text{-}1 )^{[n_{\sma{ (+)}}(k_s+\pi) - n_s]}\; ( \xi_s\,\lambda \text{-}1)^{[n_{\sma{ (-)}}(k_s+\pi) - n_s]}$, where $n_{\sma{ (+)}}(k_s+\pi)$ ($n_{\sma{ (-)}}(k_s+\pi)$) is the number of positive-$\inv$ (negative-$\inv$) bands at $k_s+\pi$. Upon proving this claim, we deduce that there are $(n_{\sma{ (+)}}(k_s+\pi) - n_s)$ number of $\;\text{-}\xi_s\;$  $\W$-eigenvalues and  $(n_{\sma{ (-)}}(k_s+\pi) - n_s)$ number of $\;+\xi_s\;$ $\W$-eigenvalues. Furthermore, we apply a result presented in   (\ref{eq:setequality1d}): the $\W$-eigenvalues can either be $\pm 1$ or form complex-conjugate pairs $\lambda,\lambda^*$. Thus, the zeros of the polynomial $R(\lambda)$ correspond to $n_s$ complex-conjugate pairs of eigenvalues. 

\noindent \emph{Proof}:\\
\noindent We define $\lambda_i = \text{-}\xi_s  \,\xi_i(k_s+\pi)\,\lambda$. By a binomial-like expansion, we express the characteristic equation as:
\bal \label{eq:characexpansion}
0 \eq \sum_{m=0}^{\noc} 2^{m} \sum_{ \{ \alpha_1,\alpha_2 \ldots \alpha_m  \} } \text{det}\; \bar{Y}^{\alpha_1 \alpha_2 \ldots \alpha_m} \;\prod_{i=1;i\neq \alpha_1,\alpha_2 \ldots \alpha_m}^{\noc} (\lambda_i \text{-}1).
\end{align} 
The sum $\sum_{ \{ \alpha_1,\alpha_2 \ldots \alpha_m  \} }$ runs over the $\noc$ choose $m$ combinations of the indices $\{ \alpha_1 \ldots \alpha_m  \}$; by $x$ choose $y$ we mean $x!/y!(x-y)!$. An example of the characteristic equation for an insulator with $4$ occupied bands is
\bal \label{eq:example4}
&0 = (\lambda_1 - 1)(\lambda_2 - 1)(\lambda_3 - 1)(\lambda_4 - 1) + \bigg[ 2\,\bar{Y}^1 (\lambda_2 - 1)(\lambda_3 - 1)(\lambda_4 - 1)  + 8\;\text{det}\;[\bar{Y}^{123}]\;(\lambda_4{-}1) + \text{comb.} \;{1} \bigg] \lin
&+ \bigg[ 4\;\text{det}\;[\bar{Y}^{12}]\;(\lambda_3 - 1)(\lambda_4 - 1) + \text{comb.} \;{2}\bigg] + 16\;\text{det}\;[\bar{Y}^{1234}]
\end{align}
where $[f(1234) + \text{ comb.}\, {1}]  = f(1234) + f(4123) + f(3412) + f(2341)$ and $[f(1234) + \text{comb.} \,{2}] = f(1234) + f(1324) + f(1423) + f(2314) + f(2413) + f(3412)$. Applying  the rank-minor theorem, only the first $n_s\pone$ terms  in (\ref{eq:characexpansion}) are nonzero:
\bal \label{eq:truncated}
0 = \sum_{m=0}^{n_s} 2^{m} \sum_{\{\alpha_1,\alpha_2 \ldots \alpha_m\}} \text{det}\; \bar{Y}^{\alpha_1 \alpha_2 \ldots \alpha_m} \;\prod_{i=1;i\neq \alpha_1,\alpha_2 \ldots \alpha_m}^{\noc} (\lambda_i \text{-}1).
\end{align}
Let us organize this expansion. We first consider a term in (\ref{eq:truncated}) with a particular combination of $m$ band indices given by the set $\{ \alpha_1 \ldots \alpha_m \}$ in the superscript of $\bar{Y}$. Each band index $\alpha_i$ has a corresponding $\inv$ eigenvalue $\xi_{\alpha_i}$ at $k=k_s+\pi$; in this set we may define $m_{\sma{+}}$  ($m_{{-}}$) as the number of positive (negative) $\inv$ eigenvalues in the set $\{\xi_{\alpha_1}(k_s+\pi) \ldots \xi_{\alpha_{m}}(k_s+\pi) \}$; note $m_{\sma{+}} + m_{{-}} = m$. The \emph{presence} of each band $\alpha_i$ in this set implies that a factor of $(\text{-}\xi_s  \,\xi_{\alpha_i}(k_s+\pi)\,\lambda \text{-}1)$ is \emph{absent} in the product $\prod_{i=1;i\neq \alpha_1,\alpha_2 \ldots \alpha_{m}}^{\noc} (\lambda_i \text{-}1)$. In the remainder of this proof, we use the convention that $\alpha_i$ ($\beta_i$) are positive-$\inv$ (negative-$\inv$) band indices at $\pi+k_s$. We may organize the expansion by collecting terms with the same $m_{\sma{+}}$:
\bal \label{idont}
0 = \sum_{m=0}^{n_s} 2^{m} \sum_{m_{\sma{+}}=0}^{m} \;\big(\text{-}\xi_s \lambda \text{-}1\big)^{[n_{\sma{ (+)}}(k_s+\pi) - m_{\sma{+}}]} \;\big( \xi_s \lambda \text{-}1 \big)^{[n_{\sma{ (-)}}(k_s+\pi)-m+m_{\sma{+}}]}\; S_{m_{+}, m-m_+}
\end{align}
with $S_{x,y}$ defined as the sum of all minors $\text{det}\,\bar{Y}^{\alpha_1 \ldots \alpha_{m_{\sma{+}}} \beta_1 \ldots \beta_{m_{{-}}}}$ with $m_{\sma{+}}=x$ and $m_{-}=y$:
\bal \label{eq:minorsumdefine}
S_{x,y}  = \sum_{\{\alpha_1 \ldots \alpha_{x} \,\beta_1 \ldots \beta_{y}\}} \text{det}\,\bar{Y}^{\alpha_1 \ldots \alpha_{x}\, \beta_1 \ldots \beta_{y}}.
\end{align}
By definition, there are $n_{\sma{ (+)}}(k_s+\pi)$ ($n_{\sma{ (-)}}(k_s+\pi)$) number of positive-$\inv$ (negative-$\inv$) bands at $k_s+\pi$, hence the sum $\sum_{\{\alpha_1 \ldots \alpha_{x} \,\beta_1 \ldots \beta_{y}\}}$ runs over $n_{\sma{ (+)}}(k_s+\pi)$ choose $x$ combinations of positive-$\inv$ bands, multiplied by $n_{\sma{ (-)}}(k_s+\pi)$ choose $y$ combinations of negative-$\inv$ bands. As the form of (\ref{idont}) suggests, nonzero terms in the expansion have values of $m_{\sma{+}}$ ranging from a minimum of $0$ to a maximum of $n_s$. That the expansion (\ref{eq:truncated}) includes nonzero terms with  $m_{\sma{+}}=n_s, m_{-}=0$ and nonzero terms with  $m_{\sma{+}}=0, m_{-}=n_s$  is a consequence of our construction: by definition of FBOP and $n_s$,  both $n_{\sma{ (+)}}(k_s+\pi)$ and $n_{\sma{ (-)}}(k_s+\pi) \geq n_s$. We observe that:

\noindent (i) Terms in the expansion with $m_{\sma{+}}=n_s, m_{-}=0$ are proportional to $( \text{-}\xi_s \,\lambda \text{-}1)^{[n_{\sma{ (+)}}(k_s+\pi) - n_s]}$ because $n_s$ factors of $(\text{-}\xi_s\, \lambda\, \xi_{\alpha_i}(k_s+\pi) \text{-}1)$ with $\xi_{\alpha_i}=\pone$ are removed from the product $\prod (\lambda_i \text{-}1)$. All other terms in the expansion have greater powers of $( \text{-}\xi_s \,\lambda \text{-}1)$. 

\noindent (ii) Similarly, terms in the expansion with $m_{\sma{+}}=0, m_{-}=n_s$ are proportional to $( \xi_s \,\lambda \text{-}1)^{[n_{\sma{ (-)}}(k_s+\pi) - n_s]}$ and all other terms in the expansion have greater powers of $(\xi_s\, \lambda \text{-}1)$.

\noindent (i) and (ii) imply that the common factor of all terms in the expansion is  $(\text{-}\xi_s\, \lambda \text{-}1 )^{[n_{\sma{ (+)}}(k_s+\pi) - n_s]}\; ( \xi_s\,\lambda \text{-}1)^{[n_{\sma{ (-)}}(k_s+\pi) - n_s]}$. The characteristic equation is thus expressible as 
\bal
0 = \big( \text{-}\xi_s \lambda \text{-}1 \big)^{[n_{\sma{ (+)}}(k_s+\pi) - n_s]} \;\big( \xi_s \lambda \text{-}1 \big)^{[n_{\sma{ (-)}}(k_s+\pi) - n_s]}\; R(\lambda)
\end{align}
with $R(\lambda)$ a polynomial of order ${2n_s}$ -- the claim is proven.
\end{widetext}

\section{Proof of Eq. (\ref{eq:equalityset2d}) } \label{app:2dconstraint}
 
Let $\calL^{\sma{k_y}}$ be a path  in the constant-$k_y$ interval $[(\text{-}\pi,k_y),(\pi, k_y)]$; we choose the path-ordering convention that $k_x$ increases along the path $\calL^{\sma{k_y}}$. We define (i) the time-reversed path ${\cal T}\,\calL^{\sma{k_y}}$, which sweeps the same interval with opposite path-ordering (\emph{i.e.}, decreasing $k_x$), and (ii) the $\inv$-mapped path ${\cal E} \,\calL^{\sma{k_y}}$, which sweeps the interval $[(\text{-}\pi,\text{-}k_y),(\pi, \text{-}k_y)]$ with decreasing $k_x$. In combination, ${\cal  E \,T} \calL^{\sma{k_y}}  = \calL^{\text{-}k_y}$; this observation leads us to the following identity:
\bal \label{eq:masterPS}
\W^{{\dagger}}&(\calL^{\text{-}k_y}) = \W(\,{\cal T}\,\calL^{\text{-}k_y})=B_{\pi,k_y}\,\W( \,{\cal 
E \, T} \,\calL^{\text{-}k_y})\,\dg{B}_{\text{-}\pi,k_y}\lin
\eq B_{\pi,k_y}\,\W(\calL^{k_y})\,\dg{B}_{\text{-}\pi,k_y}.
\end{align}
where $B_k$ are sewing matrices defined in Sec. \ref{sec:symmconst1D}. The first equality follows from (\ref{eq:unitaryline}), the second from (\ref{eq:Wilsonsewing}). Since $B_{\pi,k_y}=B_{\text{-}\pi,k_y}$ is unitary, we have shown that $\W$ along $\calL^{\sma{k_y}}$ is equivalent to the Hermitian adjoint of $\W$ along $\calL^{\text{-}k_y}$ by a unitary transformation, as advertised.

\section{Wilson Loop of the 2D Time-Reversal Symmetric Insulator} \label{app:trs}

The time-reversal operator is written as $T = Q\,K$, with $Q^{-1}=\dg{Q}$ and $K$ the complex-conjugation operator. On spin-half single-particle states, $T^2=-I$, which implies $Q^T =-Q$. We define a matrix that sews occupied Bloch bands at $\pm k$  through $T$: $V^{ij}_k = \bra{U^i_{\text{-}k}} T \ket{U^j_k}; \;i,j=1 \ldots \noc$. $Q^T =-Q$ implies $V^T_k = \text{-} V_{\text{-}k}$. By a similar proof as one presented in App. \ref{app:1dconstraint}, one can show that $V^{-1}=\dg{V}$, which implies $[U_{\text{-}k}^i]_{\alpha} = V^{ij*}_k \,Q_{ \alpha \beta}\,[U^j_k]_{\beta}^{*}$.  Applying this relation to a Wilson line between $\text{-}k^{\sma{(1)}}$ and $\text{-}k^{\sma{(2)}}$, which are infinitesimally apart:
\bal
& \W_{\text{-}k^{\sma{(1)}}\leftarrow \text{-}k^{\sma{(2)}}}^{ij} = \bra{U^i_{\text{-}k^{\sma{(1)}}}} U^j_{\text{-}k^{\sma{(2)}}} \rangle \lin
\eq V^{il}_{k^{\sma{(1)}}} \, Q^*_{{\alpha}{\beta}} \, [U^l_{k^{\sma{(1)}}}]_{\beta} \, V^{jm*}_{k^{\sma{(2)}}} \, Q_{{\alpha}{\delta}}\,[U^{m}_{k^{\sma{(2)}}}]^*_{\delta} \\ \label{eq:trrel}
&\imp \W_{\text{-}k^{\sma{(1)}}\leftarrow \text{-}k^{\sma{(2)}}} = V_{k^{\sma{(1)}}}\,\W_{k^{\sma{(1)}}\leftarrow k^{\sma{(2)}}}^*\,\dg{V}_{k^{\sma{(2)}}}
\end{align}
By employing (\ref{eq:finitetosmall}), the relation (\ref{eq:trrel}) is generalizable to finite-length paths, with arbitrary $k^{\sma{(1)}},k^{\sma{(2)}}$.
We recall the definitions of $\calL^{\sma{k_y}}$, ${\cal T}\,\calL^{\sma{k_y}}$, ${\cal E}\,\calL^{\sma{k_y}}$ and ${\cal E \,T}\,\calL^{\sma{k_y}}$, as detailed in App. \ref{app:2dconstraint}. \bal \label{eq:masterTRS}
\W^{{\dagger}}&(\,\calL^{\text{-}k_y}\,) = \W(\,{\cal T}\,\calL^{\text{-}k_y}\,)=V_{\pi,k_y}\,\W( \,{\cal E \,T}\,\calL^{\text{-}k_y}\,)^* \,\dg{V}_{\text{-}\pi,k_y}\lin
\eq V_{\pi,k_y}\,\W(\,\calL^{k_y}\,)^*\,\dg{V}_{\pi,k_y}.
\end{align}
The first equality follows from (\ref{eq:unitaryline}), the second from (\ref{eq:Wilsonsewing}) and the third from ${\cal E \,T}\,\calL^{\sma{k_y}} =\calL^{\text{-}k_y}$. Since $V_{\pi,k_y}$ is unitary, we have shown that $\W$ along $\calL^{\sma{k_y}}$ is equivalent to the transpose of $\W$ along $\calL^{\text{-}k_y}$ by a unitary transformation, thus proving (\ref{eq:equalityset2dtrs}).(\ref{eq:masterTRS}) may be written in a familiar form:
\bal
\W^{\text{-}1}(\calL^{\text{-}k_y}) = \Theta^{\text{-}1} \; \W(\calL^{k_y})\; \Theta
\end{align}
with $\Theta=K \dg{V}_{\pi,k_y}$, and satisfying $\Theta^2=\text{-}I$. This implies that each eigenstate of $\W_{\kyi}$ at $\kyi=\{0,\pi\}$ has a degenerate Kramer's partner.

\section{Isotropy of $W$} \label{app:isotropy}

Let us define $\{\varphi(k_x)\}$ ($\{\vartheta(k_y)\}$) as  Wannier trajectories of the Wilson loop at constant $k_x$ ($k_y$). In this Section we prove that if $\{\vartheta(k_y)\}$ has a relative winding of $W$, so will $\{\varphi(k_x)\}$. This follows because:

\noi{i} The Chern number is an obstruction to a smooth gauge in the BZ, hence both sets of trajectories, $\{\vartheta(k_y)\}$ and $\{\varphi(k_x)\}$, must exhibit the same center-of-mass winding $C_1$.

\noi{ii} Let $\Mp(K_x)$, $\Mm(K_x)$ and $\Mcc(K_x)$ respectively be the number of $\pone, \mo$ and complex-conjugate-pair eigenvalues of the Wilson loop at constant $K_x = \{0,\pi\}$; $\Mp(K_x)+\Mm(K_x)+\Mcc(K_x) = \noc$, the number of occupied bands. We define 
\bal
M_d = \text{max}\,&\big\{\;\Mm(\pi)-\Mm(0)-\Mcc(0),\lin
&\Mm(0)-\Mm(\pi)-\Mcc(\pi)\;\big\};
\end{align}
if $M_d>0$ there are $M_d$ number of Wannier trajectories that directly connect the primary site (at $K_x$) and the secondary site (at $K_x+\pi$); if $M_d \leq 0$ there are none. By applying the mapping of Sec. \ref{sec:theorem1d}, it is possible to prove by exhaustion that $M_d=N_d$, where $N_d$ is defined in Eq. (\ref{eq:defineNd}). 

Given (i) and (ii), the relations between  $W, C_1$ and $N_d$ in Tab. \ref{tablegraphtheory} imply that both sets of trajectories, $\{\vartheta(k_y)\}$ and $\{\varphi(k_x)\}$, have the same relative winding $W$.

\bibliography{TI-references-2013jul13}

\end{document}